\documentclass[12pt]{iopart}
\usepackage[dvipsnames]{xcolor}
\usepackage[colorlinks=true, linkcolor=blue, urlcolor=blue, citecolor=blue]{hyperref}
\usepackage{graphicx}
\usepackage{xspace}
\usepackage{amssymb}
\usepackage{subcaption}

\newcommand{\dcc}{P2300123-v6}

\newcommand{\blu}{\mathcal{B}^{\mathrm{L}}_{\mathrm{U}}}
\newcommand{\logblu}{\log_{10}\blu}
\newcommand{\datarelease}{\url{https://doi.org/10.5281/zenodo.8063631}\xspace}
\newcommand{\datareleasePE}{\url{https://doi.org/10.5281/zenodo.8409637}\xspace}

\begin{document}

\title[Waveform systematics for lensed GWs: Posterior overlap]{Waveform systematics in identifying strongly gravitationally 
lensed gravitational waves: Posterior overlap method}

\author{
\'Angel Garr\'on,
David Keitel
}

\address{Departament de F\'isica, Universitat de les Illes Balears, IAC3--IEEC, Crta. 
Valldemossa km 7.5, E-07122 Palma, Spain}
\eads{\mailto{angel.garron@ligo.org} \mailto{david.keitel@ligo.org}}
\vspace{10pt}
\begin{indented}
\item 11 December 2023
-- \href{https://dcc.ligo.org/\dcc}{LIGO-\dcc}
\end{indented}

\begin{abstract}
Gravitational lensing has been extensively observed for electromagnetic signals,
but not yet for gravitational waves (GWs).
Detecting lensed GWs will have many astrophysical and cosmological applications,
and becomes more feasible as the sensitivity of the ground-based detectors improves.
One of the missing ingredients to robustly identify
lensed GWs is to ensure that the statistical tests used are robust
under the choice of underlying waveform models. 
We present the first systematic study of possible waveform systematics
in identifying candidates for strongly lensed GW event pairs, focusing on 
the posterior overlap method.
To this end, we compare Bayes factors from all posteriors using different waveforms 
included in GWTC data releases from the first three observing runs (O1--O3).
We find that waveform choice yields a wide spread of Bayes factors in some cases.
However, it is likely that no event pairs from O1 to O3 
were missed due to waveform choice.
We also perform parameter estimation with additional waveforms for interesting cases,
to understand the observed differences.
We also briefly explore if computing the overlap from different runs for the same event
can be a useful metric for waveform systematics or sampler issues,
independent of the lensing scenario.
\end{abstract}


\noindent{\it This is the Accepted Manuscript version of an article accepted for publication in 
Classical and Quantum Gravity. IOP Publishing Ltd is not responsible for any errors or omissions in this version 
of the manuscript or any version derived from it. The Version of Record is available online at}
\url{https://doi.org/10.1088/1361-6382/ad0b9b}.
{\it This Accepted Manuscript is available for reuse under a CC BY-NC-ND licence after the 12 month embargo 
period provided that all the terms and conditions of the licence are adhered to.}

\section{Introduction}

Gravitational lensing is a phenomenon by which a gravitational field 
can bend the paths of light or gravitational waves (GWs) that pass near it.
It has been extensively observed in the first case,
but not yet for GWs~\cite{Hannuksela:2019kle,LIGOScientific:2021izm,LIGOScientific:2023bwz}.
A first detection of gravitational lensing of GWs
would not only be an additional test of general relativity,
but also open the door to many interesting applications, such as
cosmological measurements~\cite{Sereno:2011ty,Liao:2017ioi,Cao:2019kgn,Li:2019rns,Hannuksela:2020xor},
novel tests of general relativity~\cite{Baker:2016reh,Collett:2016dey,Fan:2016swi,Goyal:2020bkm,Ezquiaga:2020dao},
or constraints on dark matter (e.g.~\cite{Basak:2021ten}).

We focus on GW signals detected using the interferometric detectors
of Advanced LIGO in Hanford, Washington and Livingston, Louisiana \cite{LIGODet}
as well as Virgo in Italy~\cite{VirgoDet},
which are now also being joined by KAGRA~\cite{KAGRADet} in Japan.
The only GW signals detected so far are emitted by compact
binary coalescences (CBCs) of binary systems composed of black holes
or neutron stars. The detections made by the LIGO--Virgo--KAGRA collaboration (LVK)
are included in the various releases of the Gravitational-Wave Transient Catalog (GWTC)
\cite{LIGOScientific:2018mvr,LIGOScientific:2020ibl,LIGOScientific:2021usb,LIGOScientific:2021djp}.
Other groups have also published GW detections, such as the most recent
``Open Gravitational-wave Catalog'' paper~\cite{Nitz:2021zwj}
and what is often referred to as works of the ``IAS team''~\cite{Olsen:2022pin},
but their results have not been used for this work.

The expected fraction of lensed events is low
at the detector sensitivities achieved so far
\cite{Ng:2017yiu,Li:2018prc,Oguri:2018muv,Wierda:2021upe}.
But the ongoing improvement of detector sensitivity~\cite{Aasi:2013wya}
will make their detection much more likely,
with most studies agreeing on a rate of around one strong lensing detection per year at LIGO design sensitivity.
Indeed, no confident detections of lensed GWs
have been made in the observational studies performed to date
\cite{Hannuksela:2019kle,Li:2019osa,McIsaac:2019use,Dai:2020tpj,Liu:2020par,LIGOScientific:2021izm,LIGOScientific:2023bwz,O3followup}.
A few intriguing candidates for strongly lensed multiple images
have been reported and studied in the literature,
in particular the GW170104--GW170814 pair~\cite{Hannuksela:2019kle,Dai:2020tpj,Liu:2020par}
from the O2 run
and the GW191103--GW191105 and GW191230--LGW200104 pairs~\cite{LIGOScientific:2023bwz,O3followup}
from the O3b run,
though generally concluded to be unlikely to be lensed.

One of the missing ingredients for being able to robustly identify lensed GWs
is to ensure that the statistical tests used are robust under the choice
of underlying CBC waveform models.
This is a special case of the general issue usually referred to as
``waveform systematics''~\cite{Abbott:2016wiq}:
For any astrophysical inference about CBCs from GW data,
it is crucial to understand the possible systematic errors in inference results
due to approximations in the waveform models used.
Since full numerical relativity simulations~\cite{Palenzuela:2020tga}
are only available
as a reference for a sparse set of points in parameter space,
the most convenient way to study the impact of waveform systematics
is to compare the results from different models,
for example from
the effective-one-body,
phenomenological
and NRSurrogate
model families.
(See~\cite{Schmidt:2020ekt} for a review of modelling approaches
and table~\ref{table:waveforms} for detailed references.)
Parameter estimation (PE) runs for GWTC data releases have always used at least two waveforms
from independent modelling approaches,
and additional studies on events of special interest
regularly compare larger numbers of models
(see, e.g.
~\cite{LIGOScientific:2016ebw,LIGOScientific:2020ufj,Colleoni:2020tgc,Estelles:2021jnz,Mateu-Lucena:2021siq,Hannam:2021pit}).
Approaches to directly quantify waveform systematic errors in some other contexts have been recently proposed,
e.g.~\cite{Read:2023hkv,Owen:2023mid}.

No dedicated studies of waveform systematics have, however,
been conducted yet for gravitational lensing analyses,
except for some tests on individual events or event pairs
conducted in parallel with this work for~\cite{O3followup}
(see below).
As a first systematic study of such effects,
here we report on the robustness of initially identifying pairs of GW events
as promising candidates for double images of a single source,
when using different binary black hole (BBH) waveform models.
This is a practical first step,
as the ``posterior overlap'' method~\cite{Haris:2018vmn}
used for this initial identification in
the largest studies so far~\cite{Hannuksela:2019kle,LIGOScientific:2021izm,LIGOScientific:2023bwz}
is computationally cheap:
it is based on comparing the posteriors already available
from single-event PE under the standard unlensed signal hypothesis.
In contrast, joint-PE methods~\cite{McIsaac:2019use,Liu:2020par,Lo:2021nae,Janquart:2021qov},
which analyse the data from both events together,
have better discrimination power between lensed and unlensed scenarios,
but require additional more expensive PE runs,
and detailed studies of waveform systematics for those methods are left for future work.
We also do not consider here another machine-learning based method~\cite{Goyal:2021hxv}
for rapid candidate identification,
because it only uses spectrograms,
which do not depend on a waveform model,
and sky maps,
which usually only depend weakly on it.
By comparison, in~\cite{O3followup}
we have also assessed the impact of waveform choice on the posterior overlap method,
although only for two specific event pairs,
and microlensing PE has also been done with different waveforms for one event.
Here, we analyse the full GWTC data sets.

This paper is organized as follows: section~\ref{sec:stronglensingintro} gives
a brief introduction on lensing and, in particular, strong lensing of 
GW signals. Section~\ref{sec:catalogs} is dedicated to discussing the GWTC releases
to date and which waveforms and sampling packages were used for the PE samples in each.
We continue in section~\ref{sec:lensingsearches} with a summary of
the lensing analyses done so far on the events from these catalogs.
In section~\ref{sec:posterior-overlap-of-waveforms} we compare the
results of the posterior-overlap method on all pairs of GWTC events
using the various released samples produced with different waveforms.
Then, in section~\ref{sec:runswithadditionalwaveforms} we present PE results we have obtained
on a selection of interesting event pairs with additional waveforms.
Here we also pay attention to the difficulties in distinguishing
genuine waveform systematics
from the effects of using different PE samplers and possible convergence issues.
We present our conclusions in section~\ref{sec:conclusions}.
We also briefly consider, in~\ref{sec:selfoverlap},
the self-overlap of posteriors from different runs for a single event
as a diagnostic for waveform systematics and/or PE convergence.

\section{Strong lensing and the posterior-overlap method}\label{sec:stronglensingintro}
Gravitational lensing gives rise to a variety of effects, one of them being strong lensing.
This phenomenon occurs when a GW signal gets to Earth coming from multiple paths with different
lengths, which, in principle, would cause us to detect duplicated ``images'' of the same event.
However, due to the limited sky resolution of GW detectors for transient events,
in practice they are not distinguished geometrically, but only by a time delay,
while the intrinsic parameters like masses and spins inferred from each detection
should be very similar.

Hence, to identify interesting pairs of signals as candidates for strong lensing,
a ``posterior overlap''~\cite{Haris:2018vmn} is calculated.
This method compares the evidences of two competing hypotheses --
$\mathcal{H}_\mathrm{L}$ stating that two signals are strongly lensed images of the same source event
and $\mathcal{H}_\mathrm{U}$ that they come from independent sources.
A Bayes factor then gives the ratio between both evidences as
\begin{equation}
\blu=\frac{\mathcal{Z}_{\mathrm{L}}}{\mathcal{Z}_{\mathrm{U}}},
\end{equation}
where $\mathcal{Z}_{\mathrm{L}}=p({d_A, d_B}|\mathcal{H}_\mathrm{L})$ and $\mathcal{Z}_{\mathrm{U}}=p({d_A, d_B}|\mathcal{H}_\mathrm{U})$
and the data of the two events are given by $d_{A,B}$.

Under $\mathcal{H}_\mathrm{U}$, as the events are independent,
\begin{equation}
\mathcal{Z}_\mathrm{U}=p(d_A)p(d_B).
\end{equation}
On the other hand, the evidence for $\mathcal{H}_\mathrm{L}$ is
\begin{equation}
\mathcal{Z}_\mathrm{L}=\int \rmd \theta p(\theta)p(d_A|\theta)p(d_B|\theta),
\label{eq:zl}
\end{equation}
where $\theta$ refers to the parameters that should have similar
posterior distributions for both events under the lensing scenario.
These parameters are the masses of the two merging objects $m_{1,2}$ (in the detector frame, i.e. redshifted
by a factor $z+1$);
the right ascension and the sine of the declination, $\alpha$ and $\sin (\delta)$, respectively;
the dimensionless spin magnitudes of the two objects $a_{1,2}$;
the cosines $\cos (\theta_{1,2})$ of the tilt angles of the spins;
and the cosine $\cos (\iota)$ of the orbital inclination angle.

Using Bayes' theorem
\mbox{$p(d|\theta)=p(\theta|d)p(d)/p(\theta)$}
in~\eref{eq:zl} gives
\begin{equation}
\mathcal{Z}_\mathrm{L}=p(d_A)p(d_B)\int \rmd \theta \frac{p(\theta|d_A)p(\theta|d_B)}{p(\theta)}.
\end{equation}
Finally, we arrive at
\begin{equation}
\blu=\int \rmd \theta \frac{p(\theta|d_A)p(\theta|d_B)}{p(\theta)}.
\label{eq:bayesfactor}
\end{equation}
This Bayes factor will be large when the posterior distributions coming from the two data sets $d_A$ and 
$d_B$ have significant overlap and ideally are peaked at roughly the same points in parameter space,
but only if the posteriors are not too broad compared to the prior $p(\theta)$.

We notice that this Bayes factor is factorized between the posteriors of the two events.
Therefore, it can be implemented on posterior samples previously obtained
in single-event inference runs.
In this derivation, the prior is assumed to have been the same for both analyses.
This has often not been strictly the case for GWTC sample releases,
where e.g. mass prior ranges have been reduced for each event following preliminary estimates
to save computing power.
However, as long as the priors are uniform, which they were in the GWTC setups,
a larger range covering the full set of events can be assumed,
and the prior can be factored out of the integral
as a constant pre-factor
that does not influence the performance of $\blu$ as a ranking statistic.
We have assumed the prior ranges given in table~\ref{table:prior-ranges}.
For aligned-spin waveforms we use a smaller prior volume,
not including $\cos (\theta_{1,2})$.
For the special case of runs with zero spin priors 
we further remove $a_1$ and $a_2$ from the prior volume.

\Table{\label{table:prior-ranges}
Prior ranges used for the computation of the integral~\eref{eq:bayesfactor}.
}
\br
parameter & range \\
\mr
$m_{1,2}$ & [$2,200]\,M_\odot$\\
$\alpha$ & [$0,2\pi$]\\
$\sin{\left ( \delta \right )}$ & [$-1,1$]\\
$a_{1,2}$ & [$0,1$]\\
$\cos{\left (\theta_{1,2}\right )}$ & [$-1,1$]\\
$\cos{\left (\iota \right )}$ & [$-1,1$]\\
\br
\end{tabular}
\end{indented}
\end{table}

For our calculations of $\blu$ in this study,
we have followed~\cite{Haris:2018vmn}, but with some
slight modifications as also done in~\cite{LIGOScientific:2021izm,LIGOScientific:2023bwz}.
We split the computation of the integral in~\eref{eq:bayesfactor} into
one involving just the intrinsic parameters and another one
involving the sky parameters.
For the intrinsic parameters,
we use Gaussian kernel density estimators as in~\cite{Haris:2018vmn},
while for the paremeters giving the position in the sky we generate a two-dimensional histogram
from the samples of each event and count the number of samples that fall into each bin, 
then multiply these.

Other types of overlap estimators have also been used in the literature,
for example Bhattacharyya coefficients in~\cite{Caliskan:2022wbh,Goyal2023}.
Based on that coefficient for the mass parameters,
a sky overlap integral and a Gaussian approximation to phase offsets,
\cite{Caliskan:2022wbh} have found high probabilities of false alarms
in the sense of two unrelated CBCs producing posteriors
with large overlap, similarly to a lensed pair,
especially as catalog sizes grow with improving detector sensitivity.
Similar conclusions are expected to hold for the posterior overlap method
as discussed here,
hence the importance of following up a large number of the higher-ranked
candidates with joint-PE methods~\cite{McIsaac:2019use,Liu:2020par,Lo:2021nae,Janquart:2021qov}
and other investigation strategies~\cite{O3followup}.

In all the strong lensing searches performed so far (see section~\ref{sec:lensingsearches}),
the posterior overlap has been combined with another ranking,
which takes into account the time delay between the events constituting the pair.
This time delay should be relatively short for galaxy lenses
(typically hours or days~\cite{Ng:2017yiu, Li:2018prc, Oguri:2018muv}),
and hence such pairs are typically preferred.
However, much longer delays are possible for galaxy cluster lenses~\cite{Smith:2017mqu, Smith:2018gle, Robertson:2020mfh}.
For this reason, we only consider the $\blu$ and do not make any model assumptions on time delay throughout this paper.

\section{GWTC event catalogs and data releases}\label{sec:catalogs}

We will consider pairs of events from the catalogs published by
the LIGO--Virgo and LVK collaborations from the observing runs from O1 to O3.
In this section we will introduce for each run 
how many events were detected and
which waveforms were used for their PE. We also summarize  
this information in table~\ref{table:waveforms},
along with the references for each waveform.
These waveforms are all implemented in LALSimulation~\cite{lalsuite}.
We only consider results with BBH waveforms excluding the samples 
in the data releases obtained with tidal models (for binaries including neutron stars).
However, we do not filter by astrophysical source type probabilities,
which means that we keep some events likely to belong
to another event class, as long as some results
using BBH waveforms were released for them.
In our analyses we include events
that appear in any of the LVK data releases,
even if some of them are no longer included
in the next catalog version
due to reanalyses or changed inclusion criteria.
(E.g. for GWTC-1 and GWTC-2 the criterion was by false-alarm rate only 
but for GWTC-2.1 and GWTC-3 it was changed to a threshold on astrophysical probability).

Besides different waveforms, it is crucial to note that various different inference packages were also used
for these data releases.
First, LALInference~\cite{Veitch:2014wba}, a C-based software that has been used since 
2014 for PE of CBCs and implements both Markov-Chain Monte Carlo and nested sampling.
A newer option for Bayesian sampling is the Python package bilby~\cite{Ashton:2018jfp}, 
which implements a number of different samplers.
The dynesty sampler~\cite{Speagle:2019ivv} is widely used for most standard runs.
Finally, RIFT~\cite{Pankow:2015cra,Lange:2017wki,Wysocki:2019grj}
is a highly-parallelizable package that is typically used for more 
expensive waveforms like SEOBNRv4PHM due to its capability of performing rapid inference.
But it makes several approximations, such as not using calibration uncertainty marginalization.

\fulltable{\label{table:waveforms}
Summary of the waveforms used for the analyses present in the data releases of GWTC-1, 2, 2.1 and 3.
The data release columns show the number of events
for which that data release contains posterior contain samples
obtained with a given waveform.
}
\br
&& \centre{2}{Properties} & \centre{4}{data release}\\
\ns
waveform & refs & \crule{2} & \crule{4}\\
&& HM & Precessing & GWTC-1 & GWTC-2 & GWTC-2.1 & GWTC-3\\
\mr
IMRPhenomD$^{*}$ &\cite{Husa_2016IMRPhenomD,Khan_2016IMRPhenomD}& no & no & \00 & 36 & \00 & \00\\
IMRPhenomPv2 &\cite{Husa_2016IMRPhenomD,Khan_2016IMRPhenomD,Hannam:2013oca}& no & yes & 10 & 36 & \00 & \00\\
IMRPhenomHM$^{\dagger}$ &\cite{London:2017bcn}& yes & no & \00 & \02 & \00 & \00\\
IMRPhenomPv3HM$^{\dagger}$ &\cite{Khan:2018fmp,Khan:2019kot}& yes & yes & \00 & \03 & \00 & \00\\
IMRPhenomXPHM &\cite{Pratten:2020ceb}& yes & yes & \00 & \00 & 53 & 34\\
NRSur7dq4 &\cite{Varma:2019csw}& yes & yes & \00 & 22 & \00 & \00\\
SEOBNRv3 &\cite{Taracchini:2013rva,Pan:2013rra,Babak:2016tgq}& no & yes & 10 & \00 & \00 & \00\\
SEOBNRv4\_ROM$^{\dagger}$ &\cite{Bohe:2016gbl}& no & no & \00 & \02 & \00 & \00\\
SEOBNRv4HM\_ROM$^{\dagger}$ &\cite{Cotesta:2018fcv,Cotesta:2020qhw}& yes & no & \00 & \02 & \00 & \00\\
SEOBNRv4P &\cite{Pan:2013rra,Babak:2016tgq,Ossokine:2020kjp}& no & yes & \00 & 35 & \00 & \00\\
SEOBNRv4PHM &\cite{Ossokine:2020kjp}& yes & yes & \00 & 31 & 39 & 36\\
\br 
\end{tabular*}
$^{*}$Aligned-spin waveform, but runs set to zero spins.

$^{\dagger}$Listed here for completeness, but no pairs analyzed due to
low number of available results.
\end{table}

\subsection{GWTC-1}
This catalog~\cite{LIGOScientific:2018mvr} and associated data release~\cite{datarelease:gwtc1}
include 11 events from O1 and O2.
These were found by the pipelines PyCBC~\cite{pycbc,Usman:2015kfa}, 
GstLAL~\cite{Sachdev:2019vvd,2017PhRvD..95d2001M} and cWB~\cite{Klimenko:2015ypf}.
The PE analyses were performed
with the LALInference sampler~\cite{Veitch:2014wba}
and the waveforms IMRPhenomPv2 and SEOBNRv3,
with the exception of the binary neutron star event GW170817~\cite{gw170817}.
We exclude this event from our analyses.

\subsection{GWTC-2}\label{sec:gwtc2}
The data release~\cite{datarelease:gwtc2} associated with the second GWTC catalog~\cite{LIGOScientific:2020ibl}
includes posterior samples for 39 events detected during O3a.
These were found by the pipelines cWB, GstLAL, PyCBC and 
a focused version PyCBC-BBH~\cite{Nitz:2020oeq}.
These results were obtained with different waveforms depending on 
the event, selecting for the BBH subset from the following: IMRPhenomD, IMRPhenomHM, IMRPhenomPv2,
IMRPhenomPv3HM, NRSur7dq4, SEOBNRv4(\_ROM), SEOBNRv4HM(\_ROM) SEOBNRv4P, and SEOBNRv4PHM.
The runs with IMRPhenomD were set to zero spins
for purposes of population analyses~\cite{LIGOScientific:2020kqk},
even though the waveform does
support aligned spins.
Three different samplers were used: LALInference, bilby and RIFT.
The only event excluded from our analyses is the second binary neutron star detection 
GW190425~\cite{LIGOScientific:2020aai}.
However, no single waveform has results for more than 36 events,
because, for example GW190521~\cite{LIGOScientific:2020iuh,LIGOScientific:2020ufj}
and GW190814~\cite{LIGOScientific:2020zkf} were treated specially by the collaboration.

\subsection{GWTC-2.1}
The GWTC-2.1 paper~\cite{LIGOScientific:2021usb} presented a reanalysis of data until O3a,
including updated posterior samples for all events from O1--O2 (the same as in GWTC-1)
and new search results for O3a.
These were found by the pipelines MBTA~\cite{Aubin:2020goo}, GstLAL, PyCBC and PyCBC-BBH.
For that run, the GWTC-2.1 event list does not fully overlap with the one from GWTC-2:
some events from GWTC-2 no longer pass the modified thresholds of GWTC-2.1,
while new events were added.
In total, the GWTC-2.1 data release~\cite{ligo_scientific_collaboration_and_virgo_2022_6513631}
contains posterior samples for 55 events, after
dropping 3 original GWTC-2 candidates but finding 8 additional ones.
The waveforms used 
were IMRPhenomXPHM (for all events except the two binary neutron stars, obtained with bilby)
and SEOBNRv4PHM (for a subset, obtained with RIFT).

\subsection{GWTC-3}
The data release~\cite{GWTC-3-release} associated to this catalog~\cite{LIGOScientific:2021djp}
adds 36 GW events from O3b.
These were found by the pipelines cWB, GstLAL, MBTA, PyCBC and PyCBC-BBH.
Of these, the catalog paper only considers 35 to be above threshold,
with the fainter neutron star--black hole candidate GW200115\_042309~\cite{LIGOScientific:2021qlt}
also included in the data release.
The waveforms IMRPhenomXPHM (with the bilby sampler)
and SEOBNRv4PHM (with the RIFT sampler) were used
for all events, with the exception of GW200115\_042309 and GW191219\_163120 
(another marginal neutron star--black hole candidate) for IMRPhenomXPHM.

\section{Lensing searches on GWTC events}\label{sec:lensingsearches}
We summarize now the main searches for strongly lensed multiple images
that have been performed so far on the GWTC-1--3 events.

In~\cite{Hannuksela:2019kle} the first systematic search for lensed GWs
was performed on the events detected during O1 and O2.
The authors studied different lensing
effects, such as strong lensing or wave optics effects.
In particular, for the case of strong lensing, 
the posterior-overlap method was used on all event pairs
using the IMRPhenomPv2 samples from the GWTC-1 data release~\cite{datarelease:gwtc1},
finding no significant (above $2\sigma$) pairs
when compared to a simulated background of unlensed pairs
and also taking into account a time-delay prior.
The GW170104--GW170814 pair was later reconsidered by \cite{Liu:2020par,Dai:2020tpj}
as the most interesting case from these observing runs,
but is still considered unlikely to be lensed
due to expected lensing rates and its properties requiring unusual lensing geometry.

Various lensing analyses on the data from O3a were presented by the LIGO-Virgo collaboration~\cite{LIGOScientific:2021izm},
including a hierarchical search for multiple lensed images from a single source,
starting with posterior overlaps
(using the IMRPhenomPv2 samples from the GWTC-2 data release~\cite{datarelease:gwtc2}),
but also including joint PE and sub-threshold searches.
Once again, no evidence was found for lensed pairs.

The latest LVK paper on the topic~\cite{LIGOScientific:2023bwz}
analysed events from the full O3 run,
including the updated GWTC-2.1 event list for O3a
and the GWTC-3 events from O3b.
It again used posterior overlap
(with the IMRPhenomXPHM samples from the corresponding data releases
\cite{ligo_scientific_collaboration_and_virgo_2022_6513631,GWTC-3-release}),
but also a new machine-learning candidate identification technique~\cite{Goyal:2021hxv}
as well as joint PE and sub-threshold searches.
Some interesting candidates were identified for follow-up~\cite{O3followup}
but none claimed as lensed detections.

In addition, several sub-threshold searches for fainter counterparts to GWTC events
have been performed on O1--O3 data
\cite{Li:2019osa,McIsaac:2019use,Dai:2020tpj,LIGOScientific:2021izm,LIGOScientific:2023bwz},
and an alternative ranking method for such pairs was recently presented in~\cite{Goyal2023}.
We do not include in this study any of the candidates found from those searches,
although a waveform systematics check
for the LGW200104 counterpart candidate to the GW191230 event,
following the same approach as we use below in section~\ref{sec:runswithadditionalwaveforms},
is presented in~\cite{O3followup}.

\section{Posterior overlap using different waveforms from GWTC data releases}\label{sec:posterior-overlap-of-waveforms}

\begin{figure}
    \centering
    \includegraphics[width=0.48\textwidth]{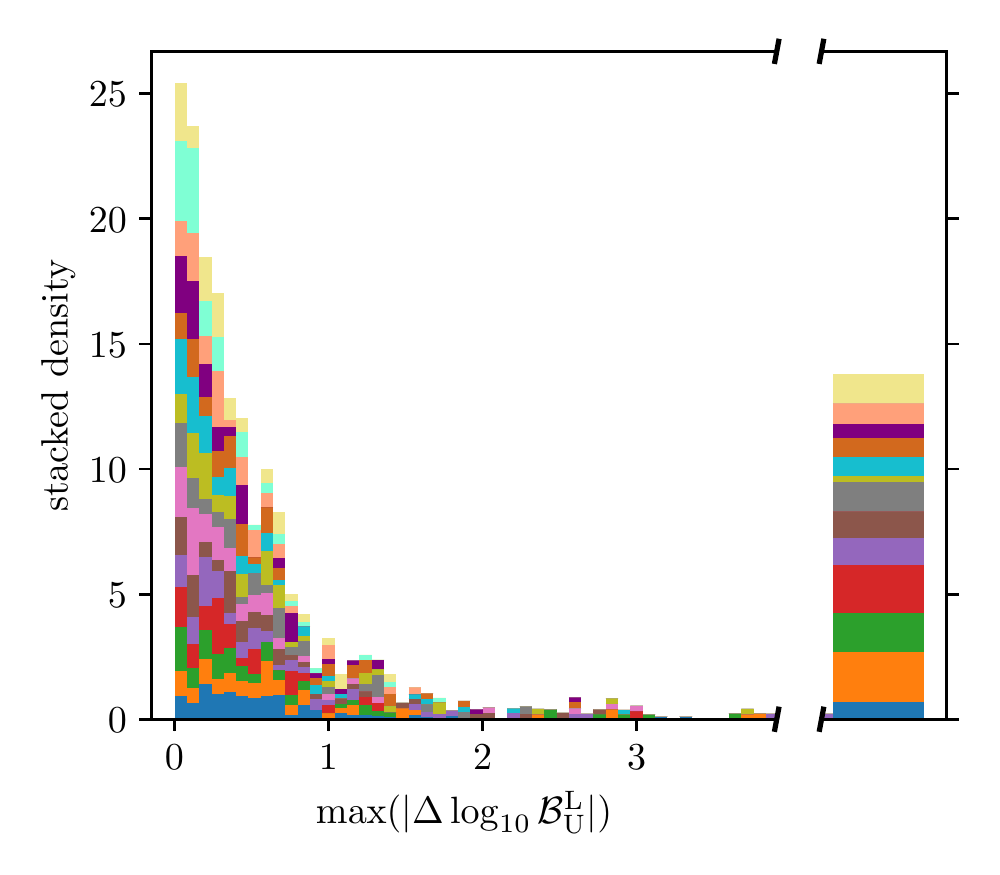}
    \includegraphics[width=0.48\textwidth]{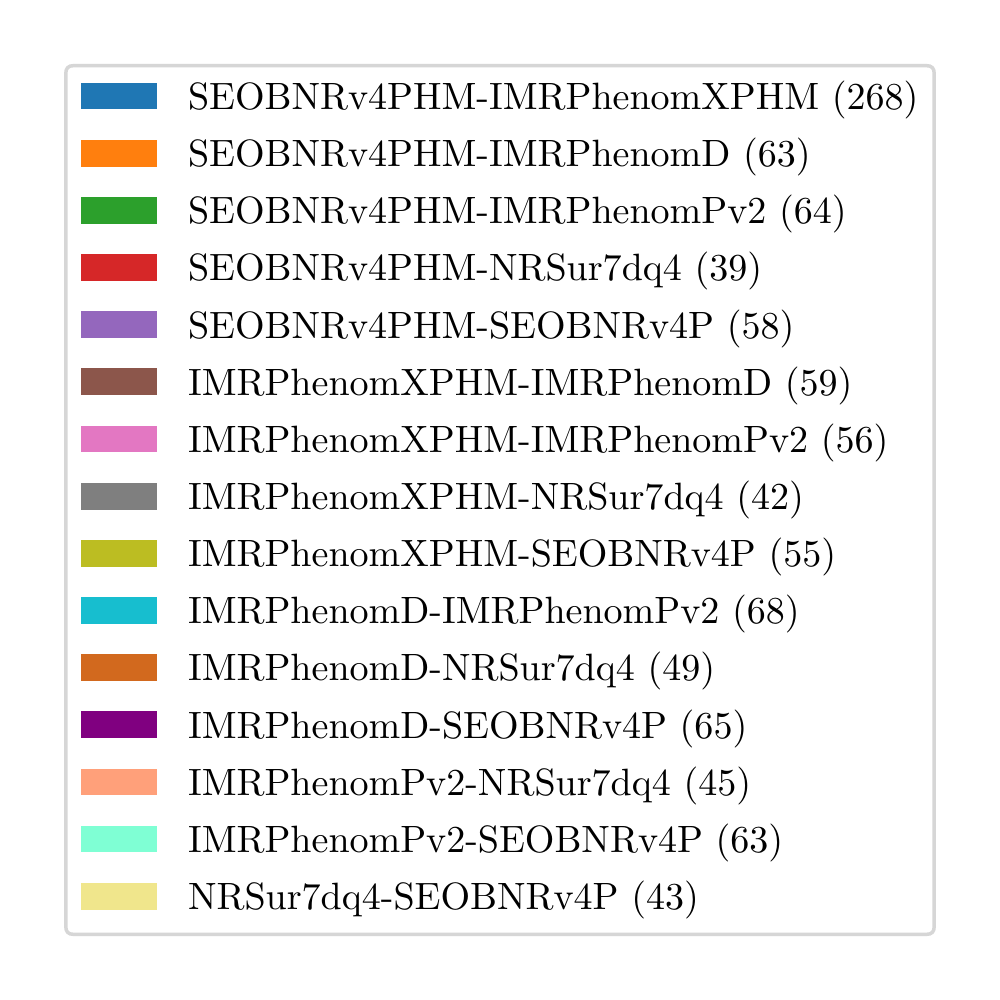}
    \caption[short]{
    \label{fig:counthistogram}
    Taking just the event pairs for which at least one waveform gives $\logblu>0$,
    we compute $|\Delta\logblu|$
    between results from different waveforms and then take the maximum
    over different catalog combinations. We show a stacked histogram
    for the density (counts normalized to an integral of one) on a given range of $\max(|\Delta\logblu|)$,
    where each color represents a different waveform combination.
    In an extra bin at the right, we collect the results for which $\max(|\Delta\logblu|)\geq 4$,
    including infinite values (where one $\blu=0$).
    The numbers in the legend indicate the number of pairs for
    each combination of waveforms.
    }
\end{figure}

Following \cite{LIGOScientific:2021izm,LIGOScientific:2023bwz},
we will focus only on event pairs from within the same observing run
(O1, O2, O3),
though as in the latter study we do include mixed O3a--O3b pairs.
Longer time delays are possible for galaxy clusters acting as lenses,
but as discussed in \cite{LIGOScientific:2021izm},
such counterparts across runs are significantly less likely.

Our main approach is to calculate the differences $\Delta\logblu$
between the log-Bayes factors computed with posteriors coming from various waveforms
and data releases\footnote{
As a sanity check, we have compared the $\blu$ results for IMRPhenomPv2 from O3a events
with those of~\cite{LIGOScientific:2021izm}, finding identical results.
On the other hand, we find small differences in $\blu$ with IMRPhenomXPHM for O3b events, compared 
with~\cite{LIGOScientific:2023bwz}, which are due to the LVK analyses having used
slightly different posterior sample files than the public release version we 
have used~\cite{GWTC-3-release}, but do not influence our conclusions.
}. 
For each event pair and pair of waveforms we also define $\max(|\Delta\logblu|)$
as the maximum over different data release combinations.
(This is necessary because different results for the same event
and the waveform SEOBNRv4PHM appear in GWTC-2 and GWTC-2.1.)
We then see which cases produced the biggest discrepancies
and perform further analyses on those.

Overall we find that the Bayes factors for a given pair can vary
significantly between different waveforms. 
Focusing on event pairs with at least one waveform giving $\logblu>0$,
we count 357 such pairs.
Of these, 79 have $\max(|\Delta\logblu|)>1$
and 27 are above 4.
The last number also includes cases
where one $\blu=0$, which means that the 
logarithmic difference is infinite.

To further illustrate which waveform combinations contribute to these discrepancies,
in figure~\ref{fig:counthistogram} we show a histogram for the number of event pairs
and waveform combinations
which give certain values of $\max |\Delta\logblu|$. 
Here we only show
the event pairs for which at least one waveform gives $\logblu>0$.

\begin{figure*}
    \includegraphics[width=0.5\textwidth]{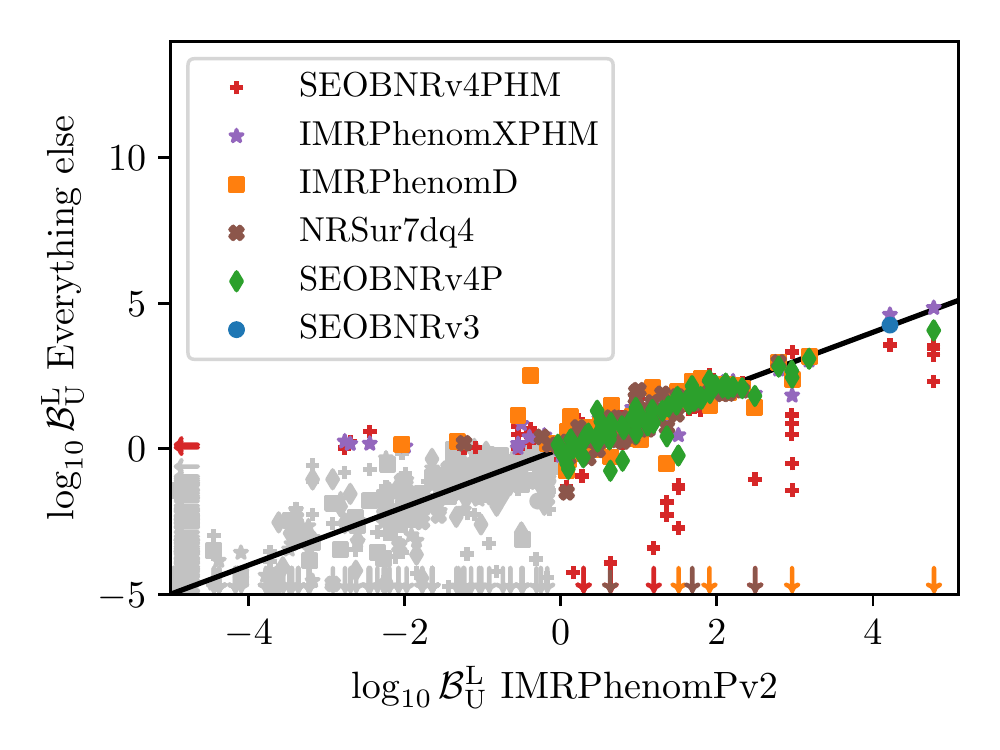}
    \includegraphics[width=0.5\textwidth]{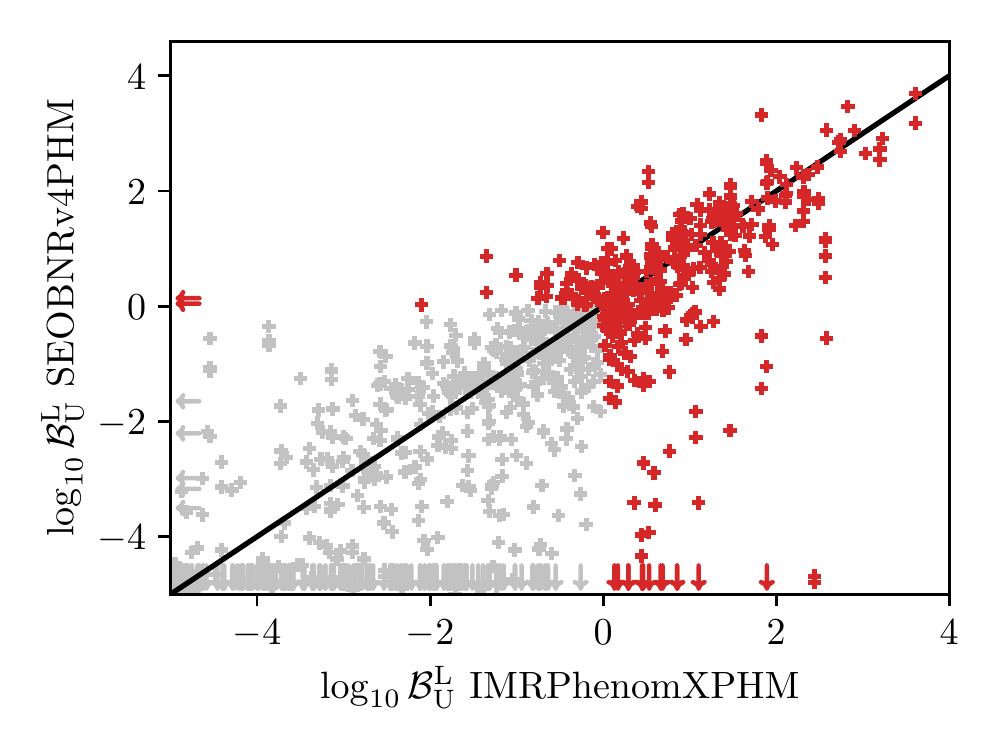}
    \caption[short]{
    \label{fig:fourpanel}
    A more detailed comparison of $\logblu$
    for event pairs, taking the posteriors computed with different waveforms.
    The left panel shows event pairs from 
    O1 to O3a (data releases GWTC-1, 2 and 2.1),
    with IMRPhenomPv2 results for each pair on the horizontal axis
    and one point drawn for comparison against any other waveform with results available for the two events
    (vertical axis).
    We highlight the points where $\logblu>0$ for at least one waveform
    and the rest are shown in gray.
    The right panel shows event pairs from O1 to O3b (data releases GWTC-2.1 and 3) comparing results from
    IMRPhenomXPHM with SEOBNRv4PHM.
    In both panels, arrows indicate cases where one of the $\logblu$ does not fit inside the plot boundaries
    and the line indicates where points should fall if the
    $\blu$ produced with different
    waveforms and included in the various catalogs were equal.
    }
\end{figure*}

In figure~\ref{fig:fourpanel} we show scatter plots comparing the 
$\logblu$ computed for different waveforms,
split into two panels going by the most common waveforms from the earlier and more recent data releases.
If we plotted all cases,
the plots would be dominated by cases where both waveforms actually
agree that $\blu$ is very low, i.e.
that the lensed scenario is strongly disfavored.
In such cases, how large $|\Delta\logblu|$ is does not make
much difference, so we will not look at these any further.

We show instead zoomed-in versions
and see clear correlations between the results from different waveforms,
although there are still quite a few outliers even for cases with at least
one waveform giving \mbox{$\logblu>0$}.
We provide a data release online\footnote{\datarelease} with the full table of event pairs and
waveform combinations corresponding to this figure,
plus all $\logblu$ values.
One noticeable pattern
is a set of outliers where other waveforms (mainly SEOBNRv4PHM) return much lower
$\blu$ than IMRPhenomPv2.
On the other hand, for cases where another waveform produces
significantly higher $\logblu$
than IMRPhenomPv2, these are mainly points just barely above $0$,
except for one case with IMRPhenomD.
Similarly, we find various outliers where
SEOBNRv4PHM gives much lower values than IMRPhenomXPHM,
and only a smaller number of less significant outliers for the opposite case.
We will now look at some of the more interesting outlier cases.

\section{Follow-up of interesting cases}\label{sec:runswithadditionalwaveforms}

Even among the pairs where at least one of the waveforms gives
$\logblu>0$, there are still too many
cases with large differences
to follow up in detail by
visual posterior inspection
and performing additional PE runs.
Hence, rather than going strictly down the list in order of $\Delta\logblu$ values,
we have focused on finding representative cases of different types of discrepancies,
such as between pairs of waveforms modelling different amounts of physics
(e.g. non-spinning/aligned-spin/precessing)
or from different modelling approaches
(IMRPhenom, EOBNR or NRSur).
Since in practical hierarchical lensing searches~\cite{Hannuksela:2019kle,LIGOScientific:2021izm,LIGOScientific:2023bwz}
only the top-ranked candidates by posterior overlap are further followed up
(e.g. with joint-PE methods),
we further focused on the pairs where at least one waveform gives
a high overlap, typically $\logblu>1$.
Also, we prioritized the followup of GW events that contribute to more than one
pair with notable $\blu$ discrepancies:
this helps us to study more pairs with the same number of PE runs,
and also to see if for those events there are intrinsic PE problems.
The latter point will also be revisited in~\ref{sec:selfoverlap}.

Other studies have also recently rerun PE on 47 GWTC events with the NRSur7dq4 waveform
using bilby~\cite{Islam:2023zzj} and on 42 events with SEOBNRv4PHM
using the machine-learning based sampler Dingo~\cite{Dax:2022pxd}.
In cases where these event selections overlap with ours, we will compare results below.

\subsection{Inference setup}
\label{sec:pe_setup}
To test if waveform modelling aspects,
such as neglecting spin precession or higher-order modes,
have a significant
impact on the PE for each event, and hence, on the overlap between pairs of events,
we perform several additional PE runs using different phenomenological waveforms
(IMRPhenomXAS, IMRPhenomXP, IMRPhenomXHM and IMRPhenomTPHM)
on each of the events from one (or more) of the selected pairs.
These additional waveforms are three variants from the same family of
frequency-domain phenomenological waveforms as IMRPhenomXPHM, as well as a
time-domain phenomenological waveform.
IMRPhenomXAS is an aligned-spin frequency-domain dominant-mode-only model~\cite{Pratten:2020fqn}.
IMRPhenomXHM is an aligned-spin frequency-domain model with higher-order modes~\cite{Garcia-Quiros:2020qpx}.
IMRPhenomXP is a precessing-spin dominant-mode-only frequency-domain model~\cite{Pratten:2020ceb}.
IMRPhenomTPHM is a precessing-spin time-domain model with higher-order modes~\cite{Estelles:2021gvs}.
Being modelled in the time domain like the SEOBNR family,
but constructed and calibrated similarly to IMRPhenomXPHM,
IMRPhenomTPHM is an ideal tool to compare the impacts of different modelling strategies.

For these runs, we use the parallel bilby package~\cite{Smith:2019ucc}
(version 1.0.1)
built on bilby~\cite{Ashton:2018jfp}
(version 1.1.5)
with the sampler dynesty~\cite{Speagle:2019ivv}
(version 1.0.1),
2000 live points and a number of
autocorrelation times of 50.
IMRPhenomX* runs typically use a single node of 128 cores
and IMRPhenomTPHM runs up to three such nodes.
Per event and waveform, we run 4 parallel chains with different random seeds
and combine the results.

For several cases, we also did additional runs with 
more expensive waveforms and newer software versions.
For NRSur7dq4~\cite{Varma:2019csw},
we used LALSimulation 5.2.0
that makes it possible to run this in a more robust setup with $f_{\min}=0$,
parallel bilby 2.0.2,
bilby 2.1.1
and dynesty 2.1.1.
For SEOBNRv5PHM~\cite{Ramos-Buades:2023ehm}
we used LALSimulation 5.2.1,
pyseobnr~\cite{Mihaylov:2023bkc} 0.2.6,
parallel bilby 2.0.2,
bilby 2.2.0
and dynesty 2.1.2.
These newer bilby and dynesty versions allow for a faster ``acceptance walk''
implementation rather than the old ``rwalk'' implementation.
It uses improved proposal distributions 
for the Markov-Chain Monte Carlo chains that are used to update live points at 
each iteration and uses the same length for all parallel chains
(determined to reach an average value of ``naccept'' accepted steps
over the run),
while the older method adapted the lengths within each chain.
We run this with 4096 live points
for NRSur7dq4 and 
2000 for SEOBNRv5PHM, naccept=60,
and on 1 or 2 nodes with 128 cores each.

For all runs,
we use the GW strain data from GWOSC~\cite{LIGOScientific:2019lzm}.
Other general settings are also matched to the GWTC analyses of each event,
including segment lengths, PSDs,
and calibration uncertainty envelopes~\cite{Sun:2020wke,Sun:2021qcg}.
For the priors, we also follow the GWTC analyses:
uniform in chirp mass and mass ratio
(with ranges changing per event
and additional constraints on component masses),
spin magnitudes (up to 0.99) and tilt angles,
inclination angle,
sky location,
and phases;
and a $d_L^2$ power law in luminosity distance (limits changing per event).
For NRSur7dq4 runs, priors are restricted
to mass ratios $m_1/m_2<6$
and total masses $\geq60M_\odot$.
For aligned-spin waveforms,
we change the spin priors to $\chi_{1,2}$ uniform in [0,0.99].
For runs with the old bilby version, as discussed e.g. in~\cite{Romero-Shaw:2020owr},
the posteriors are resampled to uniform-in-component-masses in post-processing,
while for the newer versions sampling is still in chirp mass and mass ratio,
but the correct transform from uniform-in-component-masses priors is 
directly taken into account.
So in both cases the factoring out of the prior volume in section~\ref{sec:stronglensingintro}
for $\mathcal{B}^{\mathrm{L}}_{\mathrm{U}}$ calculations
is justified.
For efficiency, the likelihood is marginalized over distance
and geocentric time of coalescence,
and those posteriors are reconstructed by interpolation
\cite{Singer:2015ema,Thrane:2018qnx,Romero-Shaw:2020owr,T1400460}.
Full settings and priors can be found in a second data release~\footnote{\datareleasePE}.

\fulltable{\label{table:followup-events}
List of event pairs we selected to follow up.
We list the two waveforms that yielded the maximum difference $|\Delta\logblu|$ 
between all waveform combinations and the $\logblu$
for each.
Full event names in the GWYYMMDD\_hhmmss format
can be found in the subsections for each pair.
}
\br
Event 1 & Event 2 & Waveform with & $\logblu$ & Waveform with & $\logblu$\\
        &         & highest $\blu$ &          & lowest  $\blu$ &         \\
\mr
GW190517 & GW190630 & IMRPhenomD & $2.52$& IMRPhenomPv2 & $-0.39$ \\
GW190503 & GW190719 & SEOBNRv4PHM & $0.69$ & NRSur7dq4 & $-\infty$\\
GW190527 & GW190719 & IMRPhenomPv2 & $1.68$ & NRSur7dq4 & $-\infty$\\
GW190527 & GW190513 & IMRPhenomPv2 & $1.36$ & SEOBNRv4PHM & $-2.28$\\
GW190527 & GW190701 & IMRPhenomD & $1.24$ & SEOBNRv4PHM & $-8.79$\\
GW190707 & GW190930 & IMRPhenomPv2 & $2.96$ & IMRPhenomD & $-44.75$\\
GW190706 & GW190719 & IMRPhenomPv2 & $2.49$ & NRSur7dq4 & $-25.73$\\
GW191103 & GW191105 & IMRPhenomXPHM & $3.03$ & SEOBNRv4PHM & \0\m$2.65$\\
\br
\end{tabular*}
\end{table}

\subsection{Results}
We discuss now the results from follow-up
studies on the event pairs listed in table~\ref{table:followup-events}.
We use the abbreviated names for the events throughout this section
and only list the full names at the start of each subsection discussing a
given pair of events.
When discussing search results for events from O3a we generally use the more recent GWTC-2.1 results for reference.
Similarly, values for their source properties are quoted from~\cite{LIGOScientific:2021usb} using
combined samples from IMRPhenomXPHM and SEOBNRv4PHM.

As discussed before, these are not all of the most extreme outliers in $\Delta\logblu$,
but rather representative cases for different situations of disagreement
between results from various waveforms.
For example,
we discuss in detail GW190517--GW190630 as a case where
the zero-spin IMRPhenomD runs from GWTC-2 produce a much higher $\blu$
than other runs.
Similar such cases where IMRPhenomD is an outlier are e.g.
GW190720--GW190930,
GW190720--GW190728
and GW190728--GW190930,
though in all of these cases the IMRPhenomD runs produced much \emph{lower}
$\blu$ than those with more complete parameter coverage,
making them less interesting to follow up.

On the other hand, we also include the GW191103--GW191105 pair,
which was highly ranked in the O3 analysis~\cite{LIGOScientific:2023bwz,O3followup},
even though its $\Delta\logblu$ are inconspicuous.

\subsubsection{GW190517--GW190630}\label{sec:GW190517--GW190630}

\Table{\label{table:blusysGW190517-GW190630}
Posterior-overlap factors for the GW190517--GW190630 pair using different waveform models in the single-event PE.
}
\br
waveform & $\logblu$ \\
\mr
IMRPhenomD$^{*}$ &  \m$2.52$\\
IMRPhenomPv2$^{*}$ &  $-0.39$\\
IMRPhenomXPHM$^{**}$ &  $-0.02$\\
IMRPhenomXAS$^{\dagger}$ &  \m$0.03$\\
IMRPhenomXHM$^{\dagger}$ &  $-0.15$\\
IMRPhenomXP$^{\dagger}$ &  \m$0.16$\\
NRSur7dq4$^{*}$ &  $-0.05$\\
SEOBNRv4P$^{*}$ &  $-0.11$\\
SEOBNRv4PHM$^{*}$ &  $-0.30$\\
SEOBNRv4PHM$^{**}$ &  \m$0.68$\\
SEOBNRv4PHM$^{\ddagger}$ &  \m$0.58$\\
SEOBNRv4PHM$^{\S}$ &  \m$0.43$\\
\br
\end{tabular}
\item[] $^{*}$Both from GWTC-2
\item[] $^{**}$Both from GWTC-2.1
\item[] $^{\dagger}$New runs for this paper
\item[] $^{\ddagger}$GW190517 from GWTC-2 and GW190630 from GWTC-2.1
\item[] $^{\S}$GW190517 from GWTC-2.1 and GW190630 from GWTC-2
\end{indented}
\end{table}

For the events GW190517\_055101 and GW190630\_185205,
we see in table~\ref{table:blusysGW190517-GW190630} 
that from the GWTC-2 posteriors obtained with IMRPhenomD
this appears as an interesting candidate with a
$\logblu$ of 2.52, while the published posteriors with other waveforms
give lower values ranging from $-0.39$ to 0.68.

GW190517 was found by all search pipelines in HLV
(Hanford--Livingston--Virgo)
data with signal-to-noise ratio
(SNR) $\approx 10$, while
GW190630 was found by GstLAL and PyCBC-BBH in LV data, but
with a higher SNR of $\approx 15.6$.
Both events are BBHs in the standard mass range
with total source-frame masses of around 60$M_\odot$
and relatively low redshifts, inferred under the unlensed hypothesis.
GW190517 has a higher effective aligned spin parameter $\chi_{\mathrm{eff}}$ of around 0.5,
while GW190630 is consistent with zero effective spin.

\begin{figure}
    \centering
    \includegraphics[width=0.3\textwidth]{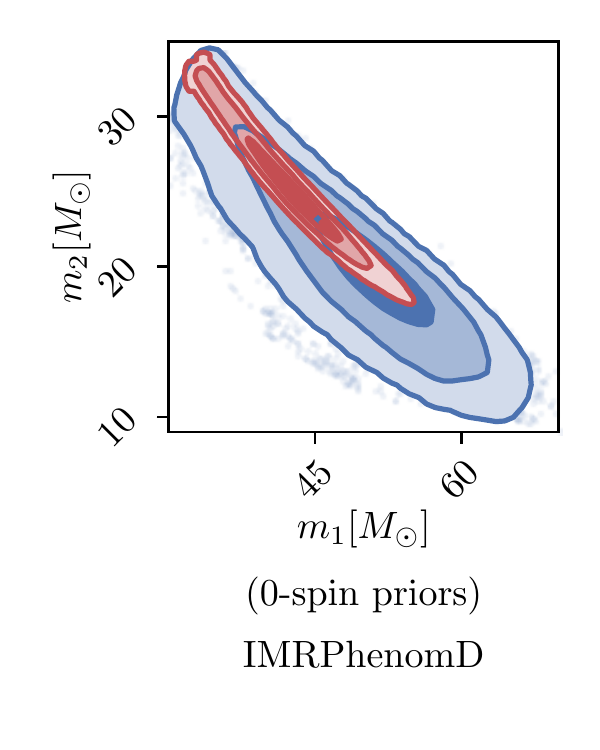}
    \includegraphics[width=0.3\textwidth]{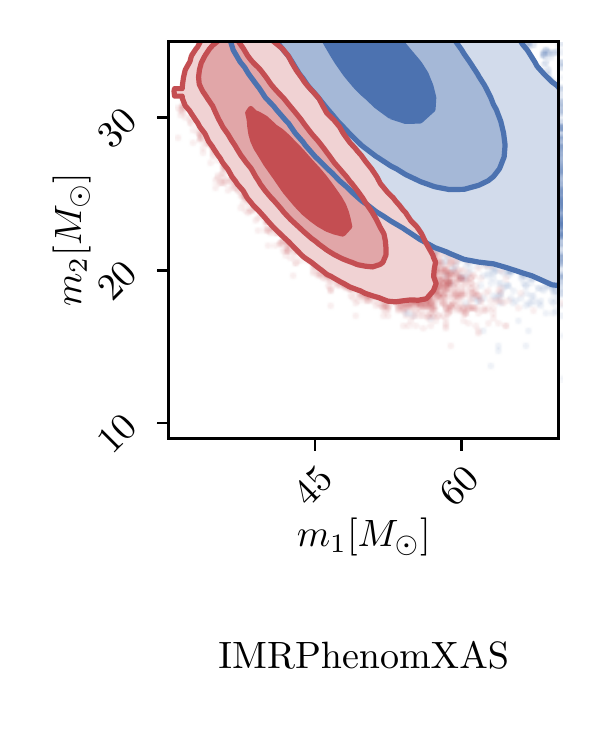}
    \includegraphics[width=0.3\textwidth]{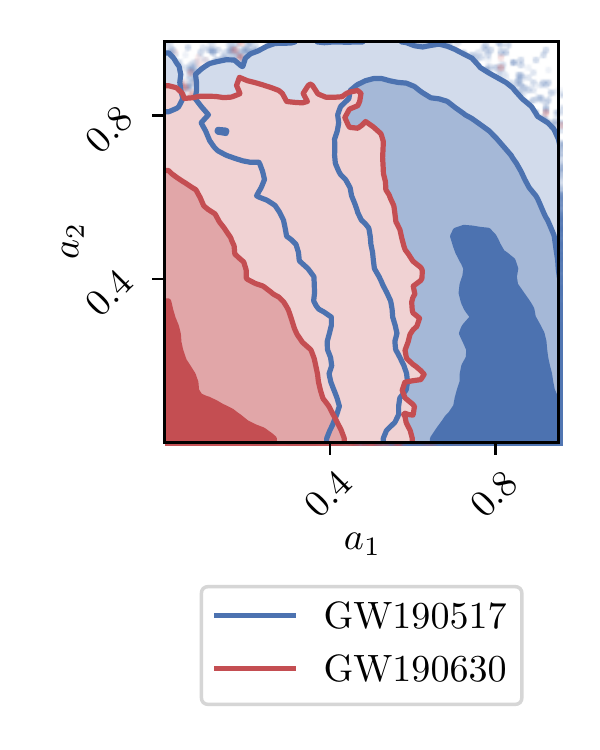}
    \caption[short]{
        Posterior overlaps for GW190517 and GW190630 with two example waveforms.
        In the first panel we show the joint $m_1$, $m_2$ distributions from 
        GWTC-2 IMRPhenomD runs, which had spins fixed to zero.
        In the other two we show the joint $m_1$, $m_2$ and
        $a_1$, $a_2$ distributions from our own runs with IMRPhenomXAS.
        We see that including spins shifts the mass posteriors, so the 
        overlap changes significantly.
    \label{fig:190517--190630}
    }
\end{figure}

We can explain the discrepancies
between the $\blu$ produced with 
PhenomD and other waveforms by the fact that, as mentioned in 
section~\ref{sec:gwtc2}, the runs with PhenomD were
fixed to zero spins.
We compare these with our own runs using
IMRPhenomXAS, which has the same physics (aligned spins and dominant mode only),
but now actually allowing non-zero spins, in figure~\ref{fig:190517--190630}.
Especially for GW190517, the lack of spins in the PhenomD run
is compensated by lower $m_2$.
Therefore, it has larger overlap with the other event.
On the other hand,
from our run with IMRPhenomXAS,
and similarly from IMRPhenomXHM and IMRPhenomXP,
we found $\blu$ that are consistent with the results
that other waveforms yield.
This indicates that the high $\blu$ from IMRPhenomD
was overestimated due to the zero-spin prior,
and that this event pair is not particularly interesting from 
the lensing point of view.

\subsubsection{GW190503--GW190719}\label{sec:GW190503--GW190719}
We further analyze the pair of events GW190503\_185404 and GW190719\_215514
because the $\blu$ found by using NRSur7dq4
was zero, while it had a value greater than 1 (positive $\logblu$) when the posteriors
from other waveforms were used -- see table~\ref{table:blusysGW190503-GW190719}.

\Table{\label{table:blusysGW190503-GW190719}
Posterior-overlap factors for the GW190503--GW190719 pair using different waveform models in the single-event PE.
}
\br
waveform & $\logblu$ \\
\mr
IMRPhenomD$^{*}$ &  $0.76$\\
IMRPhenomPv2$^{*}$ &  $0.63$\\
IMRPhenomXPHM$^{**}$ &  $0.86$\\
IMRPhenomXAS$^{\dagger}$ &  $2.12$\\
IMRPhenomXHM$^{\dagger}$ &  $2.11$\\
IMRPhenomXP$^{\dagger}$ &  $0.16$\\
IMRPhenomTPHM$^{\dagger}$ &  $0.41$\\
NRSur7dq4$^{*}$ &  $-\infty$\\
NRSur7dq4$^{\dagger}$ &  $0.59$\\
SEOBNRv4P$^{*}$ &  $0.48$\\
SEOBNRv4PHM$^{\ddagger}$ &  $0.69$\\
\br
\end{tabular}
\item[] $^{*}$Both from GWTC-2
\item[] $^{**}$Both from GWTC-2.1
\item[] $^{\dagger}$New runs for this paper
\item[] $^{\ddagger}$GW190503 from GWTC-2.1 and GW190719 from GWTC-2
\end{indented}
\end{table}

GW190503 was found by all pipelines in HLV data with SNR $\approx12$,
while GW190719 was only found by PyCBC-BBH in HL data with SNR $\approx8$.
Again quoting from the combined samples in~\cite{LIGOScientific:2021usb},
the total source-frame masses are $69.4^{+10.1}_{-8.6}$ and $57.2^{+38.4}_{-11.6} M_\odot$ 
for GW190503 and GW190719, respectively.
The posteriors in both redshift and $\chi_\mathrm{eff}$ for GW190719 tend to higher values
than for GW190503, though both overlap.

Including our own runs
we find that the aligned-spin waveforms IMRPhenomXAS and IMRPhenomXHM 
produced the highest \mbox{$\logblu\approx +2$};
while the non-spinning PhenomD runs and all runs  with precessing waveforms, except for NRSur7dq4,
agree on values between 0 and +1.
In this case the mass overlaps are relatively similar for most waveforms
(for example for IMRPhenomXAS and IMRPhenomXP the $\logblu$ computed 
with masses only is 1.8 and 1.7, respectively),
so the main difference comes from spins.
For the non-spinning PhenomD, masses are slightly biased and overlap less.
For aligned spins the mass overlap improves.
While we add more dimensions, the prior volume factor remains 1 for the aligned-spin magnitudes and
we actually obtain good overlap in spins between the
two events, with low values favored in both cases.
So for IMRPhenomXAS and IMRPhenomXHM, $\blu$ increases.
But when also allowing for in-plane spin components (precessing waveforms),
the prior volume increases again, but the posteriors become actually more
different, with higher spin magnitudes preferred for GW190719,
so the $\blu$ decreases again.

On the other hand, the difference found
between NRSur7dq4 and other precessing higher-mode waveforms
appears to not be a case of true waveform systematics.
The $\blu=0$ from the RIFT runs for NRSur7dq4 published with GWTC-2
is due to fully disjoint sky localization areas for the two events,
with GW190503 being well localized to a small region in the southern hemisphere
and GW190719 having much larger uncertainty, but constrained to the northern hemisphere.
By contrast, with other waveforms, GW190719 has a second mode in the southern hemisphere
that overlaps with GW190503.
To track down this difference,
we have performed additional reanalyses of both events using NRSur7dq4
and the latest versions of bilby (see section~\ref{sec:pe_setup}).
These reproduce the bimodal sky localization of GW190719 as with other waveforms,
and a comparable overall $\logblu\approx0.59$.
The new GW190719 run also has a broader distribution in masses than the one from RIFT,
matching better the results from phenomenological waveforms,
as will be discussed in more detail in section~\ref{sec:GW190706--GW190719}.

Hence, it appears that NRSur7dq4 was only an outlier for this event pair
(implying it to be certain to be unlensed)
due to a RIFT sampling issue.
On the other hand, aligned-spin waveforms overestimate the preference for the lensing scenario,
while bilby runs for all PHM waveforms agree on the pair
being somewhat consistent with lensing but not highly ranked.

\subsubsection{GW190527 and related events}
\label{sec:190527-and-friends}
The event GW190527\_092055 shows high overlaps and large differences 
between waveforms when compared with 
several other events: GW190719\_215514, GW170729\_185629 (although we ignore this case
due to the large time difference), GW190701\_203306,
GW190513\_205428 and others.

The event GW190527 was found by two pipelines in HL data with 
SNR $\lesssim 9$. PyCBC-BBH did not originally find it for GWTC-2,
only in the GWTC-2.1 analysis,
and with $p_{\mathrm{astro}}=0.33$,
which is lower than the $p_{\mathrm{astro}}=0.99$ given by GstLAL in GWTC-2
or $p_{\mathrm{astro}}=0.85$ in GWTC-2.1.
Its total source-frame mass is $58.1^{+18.1}_{-8.8} M_\odot$ at redshifts of 
$0.44^{+0.29}_{-0.19}$.
The event GW190513 was found by all pipelines in HLV data with
SNR $\approx 12$. Its total source-frame mass is $54.4^{+9.3}_{-6.7} M_\odot$ at redshifts of
$0.40^{+0.14}_{-0.13}$. The event GW190701 was found by all pipelines in HLV data with SNR $\approx 12$.
(The two PyCBC analyses only found it in GWTC-2.1 and not in GWTC-2,
where one of the differences is that 
the PyCBC search did not use the Virgo detector in GWTC-2.)
Its total source-frame mass is $94.3^{+12.0}_{-9.5} M_\odot$ at redshifts of
$0.38^{+0.11}_{-0.12}$. The event GW190719 has already been described in section~\ref{sec:GW190503--GW190719}.
For all of these events, $\chi_\mathrm{eff}$ is consistent with zero.

When computing $\logblu$ for the pair GW190527--GW190719,
all results (see table~\ref{table:blusysGW190527-GW190719}) 
agree on this being an interesting candidate pair, with a value
$\approx +2$, except with NRSur7dq4. So the situation is similar as for the 
previously discussed pair.
We again did a bilby rerun with NRSur7dq4 and found $\logblu$
consistent with the other waveforms. Hence, the discrepancy was, once again,
likely due to RIFT sampling issues.

For the pairs GW190513--GW190527 and GW190527--GW190701,
see tables~\ref{table:blusysGW190513-GW190527} and~\ref{table:blusysGW190527-GW190701}.
Especially for the first pair, there are again some
differences between aligned-spin and precessing models,
similar to sections~\ref{sec:GW190517--GW190630} and~\ref{sec:GW190503--GW190719},
but we do not focus on these here again.
But considering the precessing higher-mode models, for both pairs NRSur7dq4 agrees
with IMRPhenomXPHM and IMRPhenomTPHM,
while the results with SEOBNRv4PHM are outliers, yielding negative $\logblu$.
This is mostly due to the SEOBNRv4PHM mass posteriors for GW190527 being significantly narrower,
and also some differences in sky localization.
We have performed PE with bilby on the events included in 
these two pairs with the newer and faster waveform SEOBNRv5PHM
and found a $\logblu$ consistent 
with other waveforms like IMRPhenomXPHM.
Like in the previous case, we concude that the differences were related to RIFT convergence issues.

Comparing with~\cite{Islam:2023zzj}, our NRSur7dq4 results for GW190527 are consistent
but they do not include GW190719. 
They also provide NRSur7dq4 for GW190513 and GW190701 which are consistent 
with the LVK results.
Also~\cite{Dax:2022pxd} presents SEOBNRv4PHM results for GW190527 and GW190719,
finding them consistent with IMRPhenomXPHM,
while GW190513 and GW190701 are not included due to low sampling efficiency.

\Table{\label{table:blusysGW190527-GW190719}
Posterior-overlap factors for the GW190527--GW190719 pair using different waveform models in the single-event PE.
}
\br
waveform & $\logblu$ \\
\mr
IMRPhenomD$^{*}$ &  $2.30$\\
IMRPhenomPv2$^{*}$ &  $1.68$\\
IMRPhenomXPHM$^{**}$ &  $1.94$\\
IMRPhenomXAS$^{\dagger}$ &  $2.37$\\
IMRPhenomXHM$^{\dagger}$ &  $2.39$\\
IMRPhenomXP$^{\dagger}$ &  $1.93$\\
IMRPhenomTPHM$^{\dagger}$ &  $2.11$\\
NRSur7dq4$^{*}$ &  $-\infty$\\
NRSur7dq4$^{\dagger}$ &  $1.99$\\
SEOBNRv4P$^{*}$ &  $2.15$\\
SEOBNRv4PHM$^{\ddagger}$ &  $2.36$\\
SEOBNRv5PHM$^{\dagger}$ &  $1.98$\\
\br
\end{tabular}
\item[] $^{*}$Both from GWTC-2
\item[] $^{**}$Both from GWTC-2.1
\item[] $^{\dagger}$New runs for this paper
\item[] $^{\ddagger}$GW190527 from GWTC-2.1 and GW190719 from GWTC-2
\end{indented}
\end{table}

\Table{\label{table:blusysGW190513-GW190527}
Posterior-overlap factors for the GW190513--GW190527 pair using different waveform models in the single-event PE.
}
\br
waveform & $\logblu$ \\
\mr
IMRPhenomD$^{*}$ &  $-0.52$\\
IMRPhenomPv2$^{*}$ &  \m$1.36$\\
IMRPhenomXPHM$^{**}$ &  \m$1.07$\\
IMRPhenomXAS$^{\dagger}$ &  \m$2.41$\\
IMRPhenomXHM$^{\dagger}$ &  \m$2.37$\\
IMRPhenomXP$^{\dagger}$ &  \m$0.98$\\
IMRPhenomTPHM$^{\dagger}$ &  \m$0.85$\\
NRSur7dq4$^{*}$ &  \m$0.74$\\
SEOBNRv4P$^{*}$ &  \m$0.42$\\
SEOBNRv4PHM$^{**}$ &  $-2.28$\\
SEOBNRv4PHM$^{\ddagger}$ &  $-1.83$\\
SEOBNRv5PHM$^{\dagger}$ &  \m$1.10$\\
\br
\end{tabular}
\item[] $^{*}$Both from GWTC-2
\item[] $^{**}$Both from GWTC-2.1
\item[] $^{\dagger}$New runs for this paper
\item[] $^{\ddagger}$GW190513 from GWTC-2 and GW190527 from GWTC-2.1
\end{indented}
\end{table}

\Table{\label{table:blusysGW190527-GW190701}
Posterior-overlap factors for the GW190527--GW190701 pair using different waveform models in the single-event PE.
}
\br
waveform & $\logblu$ \\
\mr
IMRPhenomD$^{*}$ &  \m$1.24$\\
IMRPhenomPv2$^{*}$ &  \m$1.19$\\
IMRPhenomXPHM$^{**}$ &  \m$1.10$\\
IMRPhenomXAS$^{\dagger}$ &  \m$1.63$\\
IMRPhenomXHM$^{\dagger}$ &  \m$1.81$\\
IMRPhenomXP$^{\dagger}$ &  \m$0.94$\\
IMRPhenomTPHM$^{\dagger}$ &  \m$0.77$\\
NRSur7dq4$^{*}$ &  \m$1.09$\\
SEOBNRv4P$^{*}$ &  \m$0.89$\\
SEOBNRv4PHM$^{**}$ &  $-8.79$\\
SEOBNRv4PHM$^{\ddagger}$ &  $-3.41$\\
SEOBNRv5PHM$^{\dagger}$ &  \m$1.00$\\
\br
\end{tabular}
\item[] $^{*}$Both from GWTC-2
\item[] $^{**}$Both from GWTC-2.1
\item[] $^{\dagger}$New runs for this paper
\item[] $^{\ddagger}$GW190527 from GWTC-2.1 and GW190701 from GWTC-2
\end{indented}
\end{table}

\subsubsection{GW190707--GW190930}
We analyze the pair of events GW190707\_093326 and GW190930\_133541
because IMRPhenomD is 
an outlier, giving $\logblu=-44.75$, while
all the other waveforms give results higher than zero,
as listed in table~\ref{table:blusysGW190707-GW190930}.
In addition, the $\logblu$ for SEOBNRv4PHM, while still positive,
are noticeably lower than for other waveforms.

\Table{\label{table:blusysGW190707-GW190930}
Posterior-overlap factors for the GW190707--GW190930 pair using different waveform models in the single-event PE.
}
\br
waveform & $\logblu$ \\
\mr
IMRPhenomD$^{*}$ &  $-44.75$\\
IMRPhenomPv2$^{*}$ &  \0\m$2.96$\\
IMRPhenomXPHM$^{**}$ &  \0\m$2.57$\\
IMRPhenomXAS$^{\dagger}$ &  \0\m$5.17$\\
IMRPhenomXHM$^{\dagger}$ &  \0\m$5.28$\\
IMRPhenomXP$^{\dagger}$ &  \0\m$2.85$\\
SEOBNRv4P$^{*}$ &  \0\m$2.67$\\
SEOBNRv4PHM$^{*}$ &  \0\m$1.17$\\
SEOBNRv4PHM$^{**}$ &  \0\m$0.87$\\
SEOBNRv4PHM$^{\ddagger}$ &  \0\m$0.49$\\
SEOBNRv4PHM$^{\S}$ &  \0\m$1.13$\\
SEOBNRv5PHM$^{\dagger}$ &  \0\m$2.69$\\
\br
\end{tabular}
\item[] $^{*}$Both from GWTC-2
\item[] $^{**}$Both from GWTC-2.1
\item[] $^{\dagger}$New runs for this paper
\item[] $^{\ddagger}$GW190707 from GWTC-2 and GW190930 from GWTC-2.1
\item[] $^{\S}$GW190707 from GWTC-2.1 and GW190930 from GWTC-2
\end{indented}
\end{table}

Both events were found by all pipelines in HL data. 
GW190707 was found with SNR $\approx13$, 
with $p_\mathrm{astro}=1.00$,
while GW190930 was found with SNR $\approx10$
and $p_\mathrm{astro}$ ranging from 0.72 to 1.00.
The total source-frame masses are $\approx20$ solar masses for both events, 
and both were found with low redshifts
and $\chi_\mathrm{eff}$, 
although GW190930 is slightly peaked towards positive spins.

From our own runs, the highest $\blu$ are once again
found with the aligned-spin models IMRPhenomXAS and IMRPhenomXHM,
reaching $\logblu>5$,
while IMRPhenomXP is consistent with the results found with the
PE from IMRPhenomPv2, SEOBNRv4P and IMRPhenomXPHM given by the LVK analyses.
For all Phenom waveforms, except for IMRPhenomD,
the mass posteriors overlap very well. For IMRPhenomD, the lack of
spins causes $m_1$ for GW190707 to be peaked towards higher masses
and $m_2$ towards lower masses.
The much higher $\blu$ for IMRPhenomXAS and IMRPhenomXHM can be 
explained due to the spin posteriors for both events being very
peaked towards zero, while the precessing waveforms increase prior
volume while also returning broader posteriors.

For SEOBNRv4PHM the sky overlap is slightly better, while the
mass overlap is significantly worse, since a lower $m_2$ for GW190930 is recovered.
Again, we have performed PE with bilby and SEOBNRv5PHM and found
consistent $\logblu$ when compared with IMRPhenomXPHM.
We can conclude once again that the differences were due to 
RIFT convergence issues.

\subsubsection{GW190706--GW190719}
\label{sec:GW190706--GW190719}

For the events GW190706\_222641 and GW190719\_215514
we find negative $\logblu$ for the waveforms NRSur7dq4 and SEOBNRv4PHM,
but all other waveforms give positive values, see table~\ref{table:blusysGW190706-GW190719}.
The results with SEOBNRv4PHM in this case also differ strongly when either using posteriors from 
GWTC-2.1 for one event and GWTC-2 for the other event, or using those from
GWTC-2 for both events, but are still negative in both cases.
The pair is also interesting from the original single-waveform lensing analyses,
as it is one of the highest-ranked pairs by $\blu$
in the IMRPhenomXPHM analysis of O3 pairs~\cite{LIGOScientific:2023bwz,O3followup}
and, among those highly ranked pairs,
the second most compatible with galaxy lensing
in terms of time delay and relative magnification
\cite{O3followup,More:2021kpb,Wierda:2021upe}.

\begin{figure}
    \begin{subfigure}{0.49\textwidth}
    \centering
    \caption{IMRPhenomPv2}
    \includegraphics[width=\textwidth]{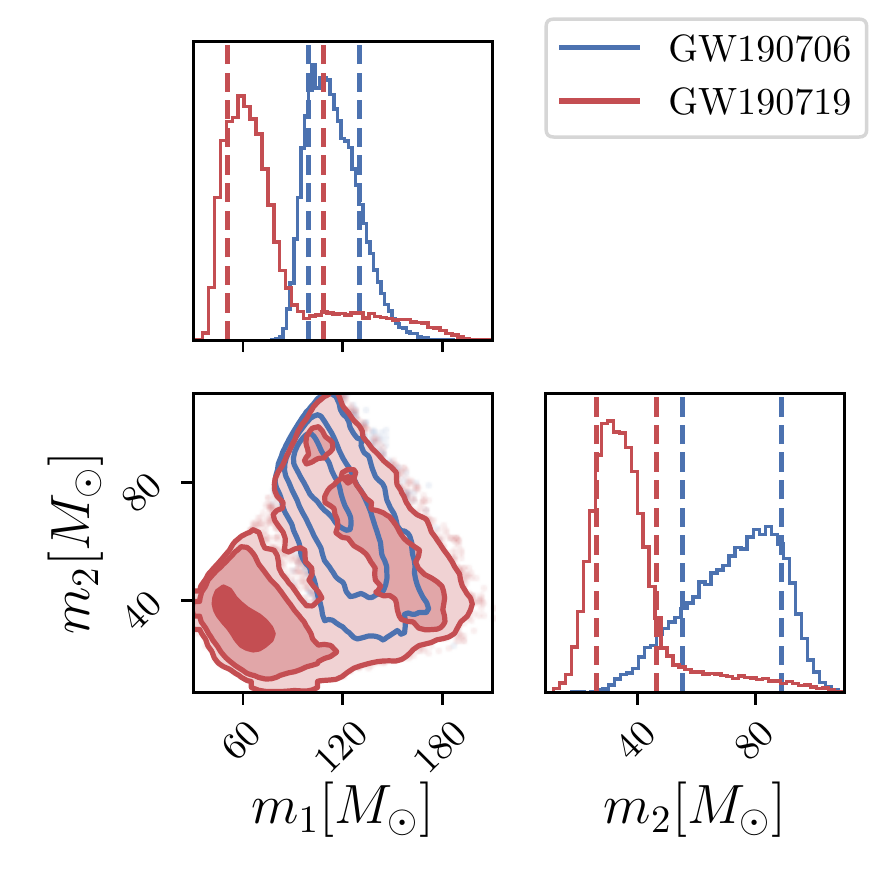}
    \end{subfigure}
    \begin{subfigure}{0.49\textwidth}
    \caption{SEOBNRv4PHM}
    \includegraphics[width=\textwidth]{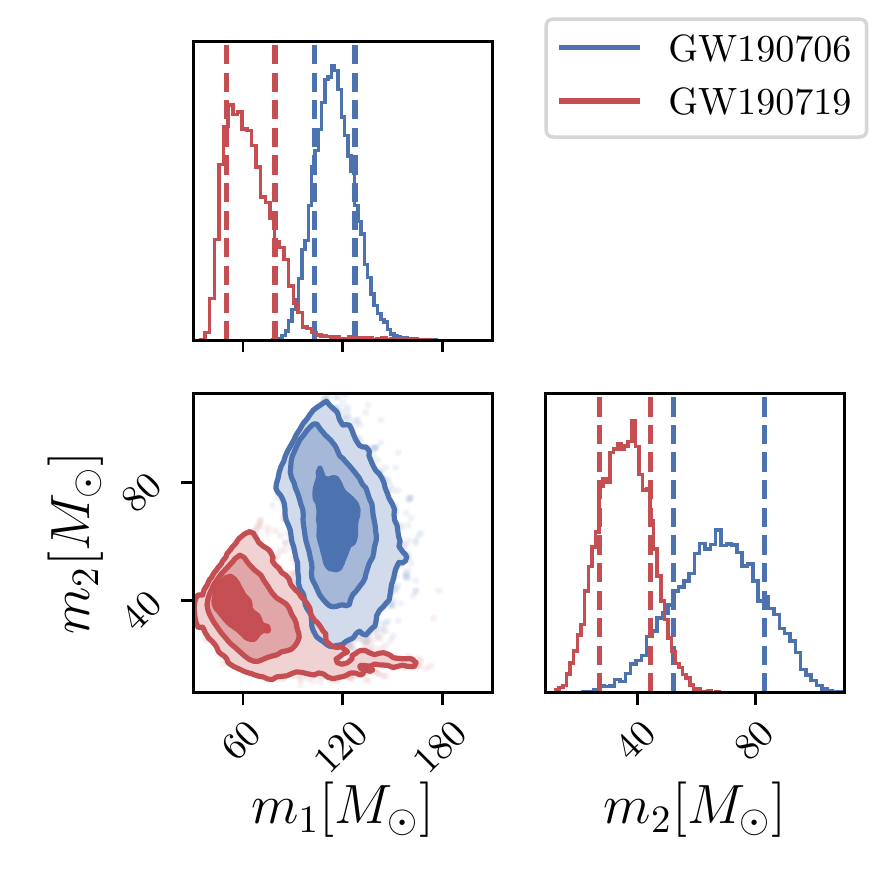}
    \end{subfigure}
    \caption[short]{
        Mass posterior distributions for the pair GW190706--GW190719 for the waveforms
        IMRPhenomPv2 (first corner plot) and SEOBNRv4PHM (GWTC-2 results, second corner plot).
    \label{fig:190706--190719}
    }
\end{figure}

The event GW190706 was found with all pipelines in HLV data
with SNR $\approx12$, while GW190719 was found only by the PyCBC-BBH
search in HL data with SNR~$\approx8$.
This makes GW190719 one of the weakest events in the catalogs,
with false-alarm rate of 1.6 per year and $p_\mathrm{astro}=0.82$ 
for GWTC-2 and 
slightly improved values of 0.6 per year and $0.92$
from GWTC-2.1.
Going by the point estimates reported in the catalog papers, we see that there 
are big differences between the masses of both events
(source-frame total masses from GWTC-2.1 of $112.6^{+27.4}_{-16.8}$ and $57.2^{+38.4}_{-11.6} M_\odot$
at quite similar redshifts of $\approx 0.6$).
The effective spins for both events are slightly positive, but still consistent with zero.
As another illustrative example, we show the detector frame posterior distributions for this pair
in figure~\ref{fig:190706--190719}.
We see that for SEOBNRv4PHM there is indeed almost no overlap,
while for IMRPhenomPv2
the overlap comes from an additional high-mass tail of the distribution for
GW190719 overlapping with the peak of the distribution for
GW190706.
The behavior of NRSur7dq4 from the LVK analyses is similar to SEOBNRv4PHM, 
while the other Phenom waveforms and SEOBNRv4P give similar results
to IMRPhenomPv2.

We also perform our own PE with the waveforms IMRPhenomXAS,
IMRPhenomXHM, IMRPhenomXP, IMRPhenomXPHM, IMRPhenomTPHM, NRSur7dq4 and SEOBNRv5PHM,
obtaining similar $\blu$ to the other Phenom or SEOBNRv4P waveforms, which are also
included in table~\ref{table:blusysGW190706-GW190719}.
We find that, in this case, the waveforms with aligned spins
yield lower $\blu$ than the precessing Phenom 
waveforms or SEOBNRv4P. This is mainly because the aligned-spin posteriors
for GW190706 extend to much higher values than for GW190719.

As for some of the other cases looked at before,
we find that our bilby NRSur7dq4 and SEOBNRv5PHM runs for GW190719 also recovered
distributions more similar to the IMRPhenom results,
with a similar additional tail of the mass distribution, which overlaps with the peak
of the distribution for GW190706 (just as shown in figure~\ref{fig:190706--190719} 
for IMRPhenomPv2).

\Table{\label{table:blusysGW190706-GW190719}
Posterior-overlap factors for the GW190706--GW190719 pair using different waveform models in the single-event PE.
}
\br
waveform & $\logblu$ \\
\mr
IMRPhenomD$^{*}$ &  \0\m$1.41$\\
IMRPhenomPv2$^{*}$ &  \0\m$2.49$\\
IMRPhenomXPHM$^{**}$ &  \0\m$1.89$\\
IMRPhenomXAS$^{\dagger}$ &  \0\m$1.40$\\
IMRPhenomXHM$^{\dagger}$ &  \0\m$1.54$\\
IMRPhenomXP$^{\dagger}$ &  \0\m$2.20$\\
IMRPhenomTPHM$^{\dagger}$ &  \0\m$2.06$\\
NRSur7dq4$^{*}$ &  $-25.73$\\
NRSur7dq4$^{\dagger}$ &  $\0\m2.25$\\
SEOBNRv4P$^{*}$ &  \0\m$1.81$\\
SEOBNRv4PHM$^{*}$ &  \0$-1.05$\\
SEOBNRv4PHM$^{\ddagger}$ &  \0$-5.77$\\
SEOBNRv5PHM$^{\dagger}$ &  \0\m$1.90$\\
\br
\end{tabular}
\item[] $^{*}$Both from GWTC-2
\item[] $^{**}$Both from GWTC-2.1
\item[] $^{\dagger}$New runs for this paper
\item[] $^{\ddagger}$GW190706 from GWTC-2.1, GW190719 from GWTC-2
\end{indented}
\end{table}

\subsubsection{GW191103--GW191105}
We also include the pair of events GW191103\_012549 and GW191105\_143521
that did not show up 
as an outlier in the $\Delta\logblu$  
distribution.
Instead, the interest in this pair comes from the combination of 
having one of the highest $\blu$ from O3 using IMRPhenomXPHM and the short time delay
between the two events,
with astrophysical priors
(at least for galaxy lenses)
contributing in favor of the lensing scenario
over background coincidences in such cases~\cite{Haris:2018vmn,More:2021kpb}.
This event pair has been studied further in~\cite{O3followup}.
As an example of a case where the results are robust under
waveform choice, we include in table~\ref{table:blusysGW191103-GW191105}
the results from both the public data releases and our own runs.
We only see some minor differences between the results obtained
with the aligned-spin and precessing waveforms; see~\cite{O3followup}
for a more detailed discussion.

\Table{\label{table:blusysGW191103-GW191105}
Posterior-overlap factors for the GW191103--GW191105 pair using different waveform models in the single-event PE.
}
\br
waveform & $\logblu$ \\
\mr
IMRPhenomXPHM$^{*}$ &  $3.03$\\
IMRPhenomXAS$^{**}$ &  $3.37$\\
IMRPhenomXHM$^{**}$ &  $3.48$\\
IMRPhenomXP$^{**}$ &  $3.08$\\
IMRPhenomTPHM$^{**}$ &  $2.70$\\
SEOBNRv4PHM$^{*}$ &  $2.65$\\
\br
\end{tabular}
\item[] $^{*}$Both from GWTC-3
\item[] $^{**}$New runs for this paper
\end{indented}
\end{table}

\section{Conclusions}
\label{sec:conclusions}

In this paper we have presented the first systematic study of possible
waveform systematics in identifying GW event pairs as candidates
for being strongly lensed copies of the same binary coalescence.
We have studied the posterior samples from the GWTC data releases.
It turns out that using posteriors obtained with different waveforms yields 
a wide spread of posterior overlap Bayes factors $\blu$ which,
in some cases, can differ by many orders of magnitude.
Even when focusing on the more promising candidates with $\blu>1$, 
we find many such cases with large differences.

However, it is likely that no event pairs from O1 to O3 were missed due to waveform choice,
because previous strong lensing studies~\cite{Hannuksela:2019kle,LIGOScientific:2021izm,LIGOScientific:2023bwz}
used the waveforms IMRPhenomPv2 and IMRPhenomXPHM
and we have found no outliers where other waveforms give a significantly higher $\blu$.

We have also followed up a representative set of outlier pairs of events,
studying their published posteriors in detail and performing
extra PE runs with the additional waveforms
IMRPhenomXAS, IMRPhenomXHM, IMRPhenomXP, IMRPhenomXPHM and IMRPhenomTPHM.
We can generally explain differences in $\blu$ between non-spinning,
aligned-spin and precessing runs from the obtained posterior distributions
and prior volumes.

Focusing on precessing higher-mode waveforms,
we found cases where apparent waveform systematics
can actually be linked to the results having been produced with different PE samplers.
GWTC results with the SEOBNRv4PHM and NRSur7dq4 waveforms
were generally obtained with RIFT.
As a cross-check,
we have also performed PE runs with NRSur7dq4 and the newer SEOBNRv5PHM
using the bilby-dynesty sampler
in a few of the cases with interesting discrepancies
and found results that better match
those obtained with IMRPhenomXPHM/IMRPhenomTPHM and bilby,
as well as earlier waveforms and LALInference.
We have thus learned that all cases of significant discrepancies 
that we have checked were due
to sampler issues rather than genuine waveform systematics.

This result is consistent with recent studies~\cite{Islam:2023zzj,Dax:2022pxd}
that also found improved agreement between PE with different waveforms
using the bilby and Dingo samplers.
It is also consistent with expectations given the detector sensitivity and observed 
population up to O3:
the current generation of precessing higher-mode waveforms
agree reasonably well in the regions of parameter space
where most events have been detected,
and by standard criteria~\cite{Lindblom:2016csg}
higher SNRs would be needed to expect noticeable systematics.
Meanwhile regions of parameter space where waveforms are less reliable 
(e.g. high spins, very unequal mass ratios)
have few detections and hence are not yet prominent in possible lensed pairs.
However, with better sensitivity and more detected events,
waveform systematics are important to consider.

A complementary way to study waveform systematics,
instead of using real events, is through injection studies.
One could either inject a simulated signal with one waveform model
and compare PE results with the same or different models,
or inject full numerical relativity waveforms
and again compare PE with various waveforms.
The main challenges for this approach are
to have a sufficiently large and representative injection set,
which covers different intrinsic parameters and relative magnifications
between lensed images, in addition to unlensed signals,
and the cost of full multi-waveform PE on many such cases.
For promising candidate pairs in the future,
studying targeted injections with similar parameters to those observed
could be an efficient approach.
An even more complex and expensive study could consist of 
injecting multiple signals into a long stretch of data and running matched-filter searches,
before doing PE, to also check for selection biases.

In~\ref{sec:selfoverlap}, we also briefly explore using the same overlap integral
as for the lensed--unlensed hypothesis test
on two posteriors from different PE runs for the same event,
as a generic diagnostic of waveform systematics and/or sampler issues.
Indeed, the outliers we have identified by this calculation are all from events
known to be challenging.
However, just as when comparing different events for the lensing test,
there is a trend for events with lower masses to have higher $\blu$,
which would need some work to take into account optimally
to develop a proper ranking statistic for runs requiring further attention.

We provide data releases of our full $\blu$ results
both for the actual lensing hypothesis test
and the single-event calculations at \datarelease
as well as our PE configurations and posterior samples
at \datareleasePE.

For the future, understanding waveform systematics in the identification of
candidates for lensed GWs
is an important ingredient for a robust detection.
Besides extending this first posterior-overlap study
with injections and different waveforms,
one could do similar studies on the strong lensing scenario
using more sensitive joint-PE methods (e.g.~\cite{Lo:2021nae,Janquart:2021qov})
as well as for the PE of single events considering micro- (e.g.~\cite{Wright:2021cbn})
or millilensing effects (e.g.~\cite{Liu:2023ikc}).

\ack
We thank the members of the LIGO--Virgo--KAGRA gravitational lensing group
for many fruitful discussions;
in particular
Jos{\'e} Ezquiaga,
Srashti Goyal,
Otto Hannuksela,
Justin Janquart,
Anupreeta More,
and Mick Wright,
We also thank
Sayantani Bera,
Marta Colleoni,
Paolo Cremonese,
H{\'e}ctor Estell{\'e}s,
Cecilio Garc{\'i}a-Quir{\'o}s,
Sascha Husa,
Maite Lucena,
Joan-Ren{\'e} M{\'e}rou,
Antoni Ramos-Buades,
Rodrigo Tenorio,
Charlie Hoy,
Colm Talbot,
and Vijay Varma
for helpful suggestions on bilby, waveform models, and computing clusters.
We also thank the referees for their helpful suggestions.

AG is supported through SOIB, the Conselleria de
Fons Europeus, Universitat i Cultura and the Conselleria de Model Econ{\`o}mic,
Turisme i Treball with funds from the Mecanisme de Recuperaci{\'o} i Resili{\`e}ncia
(PRTR, NextGenerationEU).
DK is supported by the Spanish Ministerio de Ciencia,
Innovaci{\'o}n y Universidades (ref. BEAGAL 18/00148) and cofinanced by the
Universitat de les Illes Balears.
This work was supported by the Universitat de les Illes Balears (UIB);
the Spanish Agencia Estatal de Investigaci{\'o}n (AEI) grants
PID2022-138626NB-I00, PID2019-106416GB-I00, RED2022-134204-E, RED2022-134411-T, CNS2022-135440, 
funded by MCIN/AEI/10.13039/501100011033;
the MCIN with funding from the European Union NextGenerationEU (PRTR-C17.I1);
the Comunitat Aut{\`o}noma de les Illes Balears
through the Direcci{\'o} General de Recerca, Innovaci{\'o} I Transformaci{\'o} Digital
with funds from the Tourist Stay Tax Law
(PDR2020/11 - ITS2017-006);
the Conselleria d'Economia, Hisenda i Innovaci{\'o} grant numbers 
SINCO2022/18146 and SINCO2022/6719, co-financed by the European Union
and FEDER Operational Program 2021-2027 of the Balearic Islands;
the ``ERDF A way of making Europe''.

The authors thank
the Supercomputing and Bioinnovation Center (SCBI)
of the University of M{\'a}laga
for their provision of computational resources and technical support
(www.scbi.uma.es/site)
and thankfully acknowledge the computer resources at Picasso and
the technical support provided by Barcelona Supercomputing Center (BSC)
through grants No.
AECT-2022-1-0024,
AECT-2022-2-0028,
AECT-2022-3-0024,
AECT-2023-1-0023
and AECT-2023-2-0032
from the Red Espa{\~n}ola de Supercomputaci{\'o}n (RES).
The authors are grateful for computational resources provided by the
LIGO Laboratory
and supported by
National Science Foundation Grants PHY-0757058 and PHY-0823459.

This material is based upon work supported by NSF's LIGO Laboratory
which is a major facility fully funded by the National Science Foundation.
This research has made use of data or software obtained
from the Gravitational Wave Open Science Center (gwosc.org),
a service of LIGO Laboratory, the LIGO Scientific Collaboration,
the Virgo Collaboration, and KAGRA.
LIGO Laboratory and Advanced LIGO are funded by
the United States National Science Foundation (NSF)
as well as
the Science and Technology Facilities Council (STFC) of the United Kingdom,
the Max-Planck-Society (MPS),
and the State of Niedersachsen/Germany
for support of the construction of Advanced LIGO
and construction and operation of the GEO600 detector.
Additional support for Advanced LIGO was provided by the Australian Research Council.
Virgo is funded, through the European Gravitational Observatory (EGO),
by the French Centre National de Recherche Scientifique (CNRS),
the Italian Istituto Nazionale di Fisica Nucleare (INFN)
and the Dutch Nikhef,
with contributions by institutions from
Belgium, Germany, Greece, Hungary, Ireland, Japan, Monaco, Poland, Portugal, Spain.
KAGRA is supported by
Ministry of Education, Culture, Sports, Science and Technology (MEXT),
Japan Society for the Promotion of Science (JSPS) in Japan;
National Research Foundation (NRF) and Ministry of Science and ICT (MSIT) in Korea;
Academia Sinica (AS) and National Science and Technology Council (NSTC) in Taiwan.

This paper has been assigned the document number \href{https://dcc.ligo.org/\dcc}{LIGO-\dcc}.

\appendix
\section{Posterior self-overlap as a diagnostic tool}
\label{sec:selfoverlap}

Posterior overlap as per~\eref{eq:bayesfactor} can also be computed on posteriors from the same event.
Comparing the results for different waveforms this way could be a useful diagnostic
of waveform systematics and/or sampler issues for a given event.
We perform this analysis on the same set of events and waveforms (excluding IMRPhenomD, IMRPhenomHM and IMRPhenomPv3HM), 
using the same set of parameters, as for the main lensing analysis.
The parameter choice could be revisited for this application, e.g. including the luminosity distance,
but we leave this out for a first proof of principle.
We still call the quantity computed $\blu$, even though there is no lensing hypothesis test
being performed.

\begin{figure}
    \centering
    \includegraphics[width=0.45\textwidth]{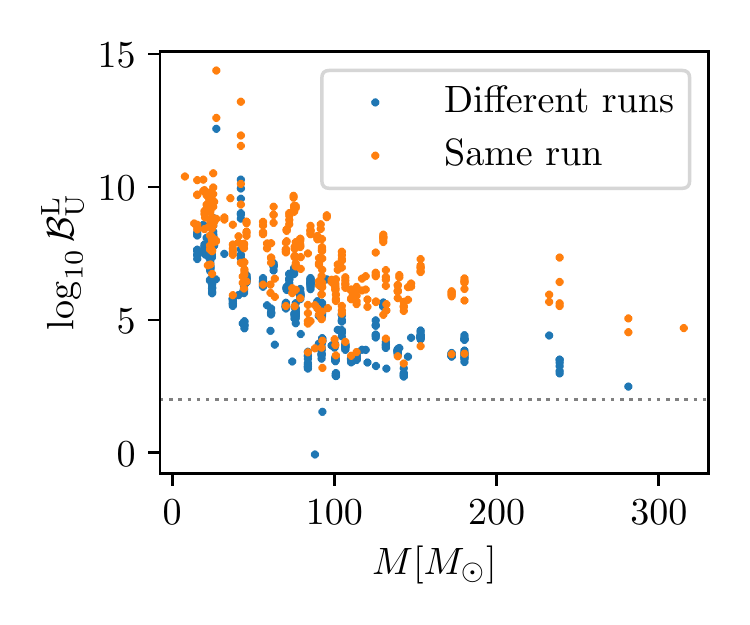}
    \includegraphics[width=0.45\textwidth]{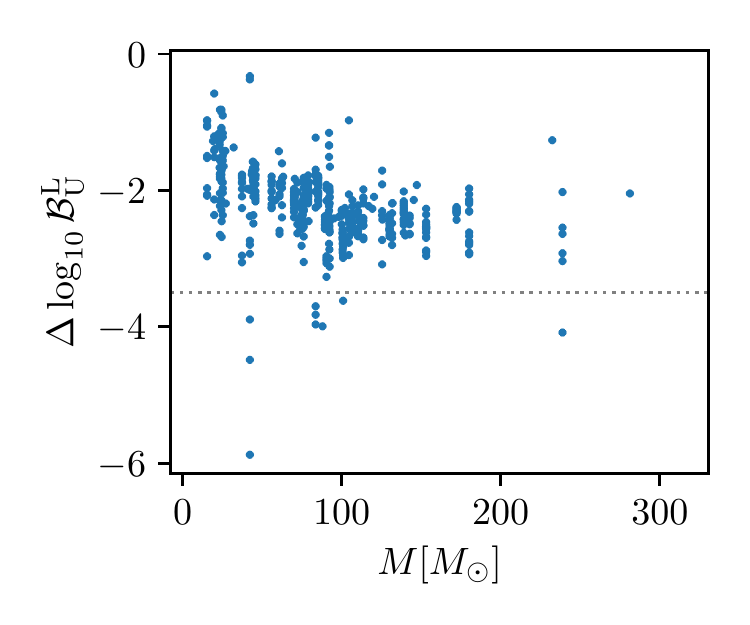}
    \caption[short]{
    \label{fig:selfoverlap}
    In the left panel we show the $\logblu$ computed from the posteriors of PE
    runs using different waveforms but the same event (blue). 
    We also show in the same panel the $\logblu$ computed from the PE coming
    from a given run with itself (orange).
    The horizontal axis is an estimate of the total detector-frame mass $M$ 
    of the event (using the median source-frame masses and redshifts from 
    GWOSC~\cite{ligo_scientific_collaboration_2022_7249086}).
    We observe some correlation that events with lower $M$ have higher $\logblu$.
    In the right panel we show 
    $\Delta \logblu = \logblu (\mathrm{run1}, \mathrm{run2})-\max ({\logblu (\mathrm{run1}), \logblu (\mathrm{run2})})$,
    which reduces this trend.
    Dotted lines indicate the ad-hoc thresholds used for discussing outliers in the text
    and in table~\ref{table:selfoverlapdifferences}.
    }
\end{figure}

In figure~\ref{fig:selfoverlap} we show the results of this test
compared against the event masses.
We observe some correlation that events with lower $M$ have higher $\logblu$.
The lower mass events have longer inspiral phases within the detector band,
giving well constrained chirp masses and hence narrower $m_{1,2}$ posteriors.
When we compute the overlap integral of two well matched posterior distributions,
then highly peaked distributions produce higher $\blu$ because the integrand
scales with their product.
We see the same trend when we just compute the overlap with~\eref{eq:bayesfactor}
from a given run with itself.

To reduce this trend, we rescale by applying the formula
\begin{equation}
\Delta \logblu = \logblu (\mathrm{run1}, \mathrm{run2})-\max \left [{\logblu (\mathrm{run1}), \logblu (\mathrm{run2})} \right ],\quad
\label{eq:rescaling}
\end{equation}
where the first term compares two different runs, e.g. with different waveforms, and the maximum is taken 
over the two overlap integrals of the posteriors from each of the two runs with itself.
This is just an ad-hoc method.
An optimal treatment of parameter dependence should start with rederiving
a single event equivalent to~\eref{eq:bayesfactor} with appropriate population
priors and taking into account the expected SNRs and posterior shapes.
The results are also shown in figure~\ref{fig:selfoverlap}
and included in the data release at \datarelease.

In the unrescaled version, the two clear outliers with low $\logblu$ at intermediate masses are
GW200322\_091133 and GW200308\_173609, both comparing the posteriors with
IMRPhenomXPHM and SEOBNRv4PHM from GWTC-3.
In~\cite{LIGOScientific:2021djp} both events are mentioned for having
modest significance, with $p_{\mathrm{astro}}=0.62$ and $0.86$, respectively, 
from only one pipeline and below $0.5$ from others, 
and significant multimodalities in their posterior distributions.

In the scaled version, the runs with the lowest $\Delta \logblu$
are listed in table~\ref{table:selfoverlapdifferences}.
These are all events already known to be challenging cases:
GW190412 was the first event with significant higher modes~\cite{LIGOScientific:2020stg}
and in-depth PE investigations were necessary to reach consistent
results across waveforms and samplers~\cite{Colleoni:2020tgc}.
The high-mass event GW190521 has also produced a variety of results
that do not always agree very well~\cite{LIGOScientific:2020iuh,LIGOScientific:2020ufj,Nitz:2020mga,Estelles:2021jnz}.
The event GW200322 is one of the outliers discussed in the previous paragraph,
which has a very broad posterior yielding quite low $\logblu$ even when comparing the same runs.
GW190527 already showed up in multiple of the pairs with large discrepancies
in our lensing analyses above (see section~\ref{sec:190527-and-friends}).
The event GW190424\_180648 was included in GWTC-2 but no longer in GWTC-2.1,
due to re-tuning of GstLAL's ranking procedure making it less likely to be astrophysical in the first place.

Hence we find that this method shows some promise in identifying events
with discrepant results from different PE runs. Further work would be required
to optimally deal with parameter scaling,
take into account other scalings, e.g. with SNR,
and compare with other metrics, like the Jensen-Shannon divergence~\cite{lin1991}.

\fulltable{\label{table:selfoverlapdifferences}
List of events and run pairs with $\Delta \logblu<-3.5$ as per~\eref{eq:rescaling}.
We include the event name,
the two runs that were used to compute the $\logblu$
(waveform name and catalog data release),
the $\logblu$ from comparing the two posteriors,
and the rescaled $\Delta \logblu$ along with which run's single-run $\logblu$ was used to to compute it.
}
\br
Event & run 1 & run 2 & $\logblu$ & $\Delta \logblu$\\
\mr
GW190412  &  SEOBNRv4PHM (GWTC-2) & SEOBNRv4PHM (GWTC-2.1) & \m$7.33$  &  $-5.88$ (run 1)\\
GW190412  &  SEOBNRv4P (GWTC-2) & SEOBNRv4PHM (GWTC-2.1) & \m$7.45$ & $-4.48$ (run 1)\\
GW190521  &  NRSur7dq4 (GWTC-2) & SEOBNRv4PHM (GWTC-2.1) & \m$3.26$ & $-4.08$ (run 2)\\
GW200322  &    IMRPhenomXPHM (GWTC-3) & SEOBNRv4PHM (GWTC-3) & $-0.06$ & $-3.99$ (run 1)\\
GW190527  &  IMRPhenomPv2 (GWTC-2) & SEOBNRv4PHM (GWTC-2.1) & \m$3.54$ & $-3.97$ (run 2)\\
GW190412 & IMRPhenomXPHM (GWTC-2.1) & SEOBNRv4PHM (GWTC-2.1) & \m$7.65$ & $-3.89$ (run 1)\\
GW190527  &  SEOBNRv4P (GWTC-2) & SEOBNRv4PHM (GWTC-2.1) & \m$3.68$ & $-3.82$ (run 2)\\
GW190527  &  NRSur7dq4 (GWTC-2) & SEOBNRv4PHM (GWTC-2.1) & \m$3.80$ & $-3.70$ (run 2)\\
GW190424  &  SEOBNRv4P (GWTC-2) & SEOBNRv4PHM (GWTC-2) & \m$2.91$ & $-3.61$ (run 1)\\
\br
\end{tabular*}
\end{table}

\section*{References}
\bibliographystyle{iopart-num-links}
\bibliography{../lensing}
\end{document}